\documentclass[preprint,superscriptaddress,nofootinbib]{revtex4-1}
\usepackage{graphicx}
\usepackage{dcolumn}
\usepackage{bm}
\usepackage{amsmath}
\usepackage{mathtools,slashed}
\usepackage{amsfonts} 
\usepackage{amssymb}
\usepackage{latexsym}
\usepackage{bbm}
\usepackage{color}
\usepackage{amssymb}
\usepackage{comment}
\usepackage{amsthm}
\usepackage{epsf}
\usepackage{epsfig}
\usepackage{caption3}
\usepackage{appendix}
\usepackage{cases}
\usepackage{subfigure}
\usepackage{multirow}
\usepackage{float}
\usepackage{makecell}

\usepackage{booktabs}
\usepackage{threeparttable}
\usepackage{mathtools}
\usepackage{diagbox}

\newcommand{\beq}{\begin {equation}}
\newcommand{\eeq}{\end   {equation}}
\newcommand{\bea}{\begin {eqnarray}}
\newcommand{\eea}{\end   {eqnarray}}
\newcommand{\baa}{\begin {array}   }
\newcommand{\eaa}{\end   {array}   }
\newcommand{\bit}{\begin {itemize} }
\newcommand{\eit}{\end   {itemize} }

\newcommand{\zboson}{{Z}}
\newcommand{\wpm}{W^{\pm}}

\newcommand{\mttwo}{M_{{T2}}} 
\newcommand{\pt}{p_{{T}}} 
\newcommand{\mt}{m_{{T}}} 
\newcommand{\HT}{H_{{T}}} 

\newcommand{\sinbma}{\text{sin}(\beta-\alpha)}
\newcommand{\tanb}{\text{tan}\beta}

\newcommand{\hboson}{{{h}}}
\newcommand{\Hboson}{{{H}}}
\newcommand{\Aboson}{{{A}}}
\newcommand{\Hpm}{{{H}}^{\pm}}
\newcommand{\mh}{M_{{{h}}}}
\newcommand{\mH}{M_{{{H}}}}
\newcommand{\mA}{M_{{{A}}}}
\newcommand{\mcH}{M_{{{H}}^{\pm}}} 

\newcommand{\gev}{\text{GeV}}

\newcommand{\threefbm}{\text{300}~\text{fb}^{-1}}

\newcommand{\cmsfourteen}{\sqrt{s}~=~\text{14 TeV}}

\begin{document}

\title{Analysis of the $ q\bar q\to Z^* \to hA \to4\tau$ process \\ within the lepton-specific 2HDM  at the LHC}
\author{Yan Ma}
\email[]{20214016005@mails.imnu.edu.cn}
\affiliation{\scriptsize College of Physics and Electronic Information,
Inner Mongolia Normal University, Hohhot 010022, PR China}

\author{A. Arhrib }
\email[]{\small a.arhrib@gmail.com}
\affiliation{\scriptsize Abdelmalek Essaadi University, Faculty of Sciences and Techniques,
B.P. 2117 Tétouan, Tanger, Morocco \& \\
 Department of Physics and Center for Theory and Computation, National Tsing Hua
University, Hsinchu, Taiwan 300}

\author{S. Moretti}
\email[]{\small s.moretti@soton.ac.uk; stefano.moretti@physics.uu.se}
\affiliation{\scriptsize School of Physics \& Astronomy, University of Southampton,
Southampton SO17 1BJ, UK \& \\
Department of Physics \& Astronomy, Uppsala University, Box 516, 75120 Uppsala, Sweden}

\author{S. Semlali}
\email[]{\small s.semlali@soton.ac.uk}
\affiliation{\scriptsize School of Physics and Astronomy, University of Southampton,
Southampton SO17 1BJ, UK \& \\
Particle Physics Department, Rutherford Appleton Laboratory, Chilton, Didcot, Oxon OX11 0QX,
United Kingdom}

\author{Y. Wang}
\email[]{\small wangyan@imnu.edu.cn}
\affiliation{\scriptsize College of Physics and Electronic Information,
Inner Mongolia Normal University, Hohhot, 010022, China
\& \\
Key Laboratory for Physics and Chemistry of Functional Materials,
Inner Mongolia Normal University, Hohhot, 010022, China}

\author{Q.S. Yan}
\email[]{\small yanqishu@ucas.ac.cn}
\affiliation{\scriptsize Center for Future High Energy Physics, Chinese Academy of Sciences,
Beijing 100049, PR China \& \\ School of Physics Sciences, University of Chinese Academy of Sciences,
Beijing 100039, PR China}

\vspace*{-1cm}
\begin{abstract}
{\footnotesize\baselineskip=15pt
We analyse light Higgs scalar and pseudoscalar associated hadro-production in the 2-Higgs Doublet Model (2HDM) Type-X (or lepton-specific) within the parameter space allowed by theoretical self-consistency requirements as well as the latest experimental constraints from the Large Hadron Collider (LHC),  precision data and $B$ physics. Over the viable regions of such a scenario, the Standard Model (SM)-like Higgs boson discovered at the LHC in 2012 is the heavier CP-even state $H$. Furthermore, in the Type-X scenario, due to  large $\tanb$, the lighter Higgs scalar $h$ and the pseudoscalar $A$  mainly decay into two $\tau$ leptons. Therefore, we concentrate on analysing the signal process $pp\to Z^{*} \to hA\to \tau^{+}\tau^{-}\tau^{+}\tau^{-}\to \ell \nu_\ell \ell \nu_\ell \tau_h \tau_h$ (where $\ell= e, \mu$ whereas $\tau_h$ represents the hadronic decay of the $\tau$) and explore the feasibility of conducting such a search at the LHC with a centre-of-mass energy of $\cmsfourteen$ and a luminosity of $L~=~\threefbm$. To suppress the huge SM background, we confine ourselves to consider the fraction of 2HDM  events with two same-sign $\tau$ leptons further decaying into same-sign leptons while the other two $\tau$ leptons decay hadronically. We find that a combination of kinematical selection and Machine Learning (ML) analysis will yield significant sensitivity to this process at the end of  LHC Run 3. 

}
\end{abstract}

\maketitle
\tableofcontents
\section{Introduction}

The LHC has played an indispensable role in improving our understanding of the microcosmic subatomic structure, particularly with the discovery of a 125 GeV scalar particle by the ATLAS and CMS collaborations in 2012~\cite{Aad:2012tfa, Chatrchyan:2012ufa}. This was the last particle predicted by the SM, and its properties, as measured by these experiments, are currently consistent with the SM prediction at the $5~\sigma$ level~\cite{ATLAS:2022vkf, CMS:2022dwd}. Investigating the Higgs sector at the LHC, where the presence of additional (pseudo)scalars is possible, remains an intriguing task, especially in new physics models with additional doublet fields. Therefore, discovering extra Higgs bosons at the LHC is one of the prime targets of Beyond the SM (BSM) signal searches.

One of the simplest extensions of the SM is the 2HDM, which contains two complex Higgs doublets~\cite{Lee:1973iz,Deshpande:1977rw,Branco:2011iw}. After Electro-Weak Symmetry Breaking (EWSB), there are three Goldstone bosons, which are "eaten" by the $\wpm$ and $\zboson$ bosons, while the remaining five degrees of freedom become physical Higgs bosons: two neutral CP-even scalars ($h$ and $H$, with $M_h \textless M_H$), one CP-odd pseudoscalar ($A$), and two charged Higgs states $H^\pm$ (with mixed CP status).

In order to be consistent with the stringent experimental constraints from Flavour Changing Neutral Currents (FCNCs) at the tree level, a $Z_2$ symmetry is  typically introduced into the Yukawa sector so that each type of fermion couples only to one of the doublets of the 2HDM~\cite{Glashow:1976nt}. Based on the $Z_2$ charge assignments of the Higgs doublets, we can define four basic scenarios, known as (Yukawa) types. In the lepton-specific scenario (or Type-X), the additional Higgs doublet $\phi_2$ couples only to the lepton sector and not to the quark sector, where the light Higgs boson mainly decays into $\tau^+\tau^-$ over a wide range of parameter space~\cite{Aoki:2009ha}.

Motivated by recent searches for $\tau^+\tau^-\tau^+\tau^-$ by the CMS experiment~\cite{CMS:2015twz, CMS:2017dmg, CMS:2019spf, CMS:2018ams}, we will study the process $pp\to Z^{*} \to hA \to \tau^+\tau^-\tau^+\tau^-$ in the 2HDM Type-X. In this scenario, the $h/A \to \tau^+\tau^-$ decay could be enhanced, resulting in a sizeable $4\tau$ final state. Therefore, it is worthwhile to examine whether this statement is robust enough after taking into account the effects of parton showering, hadronisation, heavy flavour decays and detector resolution.

The results from the muon $g-2$ experiment collected in 2021, conducted at  the Fermilab National Accelerator Laboratory (FNAL), when combined with the old ones from the Brookhaven National Laboratory (BNL), gives:
\begin{equation}
a_{\mu}^{\rm exp} = 116592061(41)\times 10^{-11},
\end{equation}
where $a_{\mu}=(g-2){\mu}/2$ is the muon anomalous magnetic moment~\cite{Albahri2021}.
The theoretical prediction can be separated into pure QED, EW and hadronic contributions. The pure QED contribution is calculated at  five loops whereas two loops are evaluated in the EW interactions. The hadronic contribution, appearing at $\mathcal{O}$(1) GeV due to Hadronic Vacuum Polarisation (HVP) and Ladronic Light-by-Light (HLbL) scattering, provides most of the uncertainty. Using a data-driven dispersive approach, one obtains $a_{\mu}^{\rm SM} = 116591810(43)\times 10^{-11}$, which yields a 4.2 $\sigma$ discrepancy between the experimental data and the theoretical prediction from the SM~\cite{Aoyama2020}:
\begin{equation}
\Delta a_{\mu} = a_{\mu}^{\rm exp} - a_{\mu}^{\rm SM} = (251\pm59) \times 10^{-11}.
\end{equation}
In 2023, the cross section measurement of the process $e^{+}e^{-} \to \pi^{+}\pi^{-}$ with the CMD-3 detector \cite{CMD-3:2023rfe} increases the tension among the data-driven dispersive evaluations of the HVP. If the theoretical calculation of the hadronic part is replaced with its value, the SM prediction will be within a 0.9$\sigma$ deviation from the FNAL experimental measurements in 2023~\cite{Ignatov2024}.
In contrast, if lattice QCD methods are used for both HVP and HLbL, as proposed by the Budapest-Marseille-Wuppertal collaboration~\cite{Borsanyi}, the experimental result is consistent with the SM prediction at 1.5$\sigma$. Thus, the muon $g-2$ SM prediction was updated in the 2025 White Paper using the lattice QCD method~\cite{Aliberti}. The latest FNAL measurement based on data taken from 2020 to 2023 gives the world average $a_{\mu}^{\rm exp} = 1165920715(145)\times 10^{-12}$, which still agrees with the SM prediction~\cite{Aguillard}.

However, one can also explain the muon anomalous  magnetic moment with new physics models, e.g., the 2HDM Type-X scenario. Since the couplings of the new Higgs bosons to the SM fermions are proportional to $\tanb$, the corresponding contributions to $\Delta a_{\mu}$ can be enhanced~\cite{Jueid2023, Iguro:2023tbk}. Although a recent measurement has already shown  the possibility of restricting the parameter space of the Type-X model in the large $\tanb$ region~\cite{CMS:CMS23007}, further research is still required.

In this paper, we assume that the heavier CP-even Higgs boson $H$ is the observed SM-like Higgs boson whose properties agree with the LHC measurements. We then perform a detailed Monte Carlo (MC) study on the signal process $pp\to Z^{*}\to hA\to 4\tau \to l\nu_ll\nu_l\tau_h\tau_h $ (with $l = e, \mu$ and $\tau_h$ representing hadronic $\tau$-decays) and relevant background processes, to examine its feasibility at the LHC.

The paper is organised as follows. In the next section, we briefly describe the 2HDM and its Yukawa scenarios, then introduce a few Benchmark Points (BPs) for our MC analysis that pass all present theoretical and experimental constraints. In the following section, we perform a detailed collider analysis of these BPs and examine the potential to discover the aforementioned signature within the 2HDM Type-X scenario. Finally, we present some conclusions.

\section{The 2HDM}
The Higgs sector of the 2HDM consists of two weak isospin doublets with hypercharge $Y = 1$. The most general  Higgs  potential for the 2HDM that complies with the ${SU(2)_L \times U(1)_Y}$ gauge structure of the EW sector of the SM has the following form  \cite{Branco:2011iw}:
\begin{eqnarray}
V(\phi_1,\phi_2) &=& m_{11}^2(\phi_1^\dagger\phi_1) +
m_{22}^2(\phi_2^\dagger\phi_2) -
[ m_{12}^2(\phi_1^\dagger\phi_2)+\text{h.c.}] ~\nonumber\\&& 
+ \frac12\lambda_1(\phi_1^\dagger\phi_1)^2 +
\frac12\lambda_2(\phi_2^\dagger\phi_2)^2 +
\lambda_3(\phi_1^\dagger\phi_1)(\phi_2^\dagger\phi_2)+ 
\lambda_4(\phi_1^\dagger\phi_2)(\phi_2^\dagger\phi_1) ~\nonumber\\ && +
\frac12\left[\lambda_5(\phi_1^\dagger\phi_2)^2 +\rm{h.c.}\right]
+~\left\{\left[\lambda_6(\phi_1^\dagger\phi_1)+\lambda_7(\phi_2^\dagger\phi_2)\right]
(\phi_1^\dagger\phi_2)+\rm{h.c.}\right\},
\label{CTHDMpot}
\end{eqnarray}
where $\phi_{1}$ and $\phi_2$ are the two Higgs doublet fields. By hermiticity of such a potential,  $\lambda_{1,2,3,4}$ as well as $m_{11,22}^2$ are real parameters, while $\lambda_{5,6,7}$ and $m_{12}^2$  can be complex, thereby enabling possible Charge and Parity (CP) violation effects in the Higgs sector. Upon enforcing the two minimisation conditions of the potential, $m^2_{11}$ and $m^2_{22}$ can be replaced by $v_{1,2}$, which are the Vacuum Expectation Values (VEV) of the Higgs doublets $\phi_{1,2}$, respectively.

Moreover, the couplings $\lambda_{1,2,3,4,5}$ can be substituted with the four physical Higgs masses (i.e., $\mh, \mH, \mA$ and $\mcH$) and the parameter $\sinbma$, where $\alpha$ and $\beta$ are, respectively, the mixing angles between the CP-even states and the angle related to the VEV values, i.e., $\tanb=v_1/v_2$. Thus, independent input parameters can be taken as $\mh, \mH, \mA$ and $\mcH$, $\lambda_6$, $\lambda_7$, $\sinbma$, $\tanb$ and $m^2_{12}$.

If both Higgs doublet fields of the general 2HDM couple to all fermions, the ensuing scenario can induce FCNCs in the Yukawa sector at tree level.  As intimated, to remedy this, a $Z_2$ symmetry is imposed on the Lagrangian such that each fermion type interacts with only one of the Higgs doublets \cite{Glashow:1976nt}.  As a consequence, there are four possible types of 2HDM, namely Type-I, Type-II, Type-X (or lepton-specific) and Type-Y (or flipped). However, such a symmetry is explicitly broken by the quartic couplings $\lambda_{6,7}$ and softly broken by the (squared) mass term $m^2_{12}$. 

In what follows, we shall consider a CP-conserving (i.e., $m^2_{12}$ and $\lambda_5$ are real) 2HDM Type-X and assume $\lambda_{6} = \lambda_{7} = 0$ to forbid the explicit breaking of $Z_2$, while also taking $m^2_{12}$ to be generally small, thereby preventing large FCNCs at tree level, which are incompatible with experimental data.

In general, the couplings of the neutral and charged Higgs bosons to fermions can be described by the Yukawa Lagrangian~\cite{Branco:2011iw}

\begin{eqnarray}
- {\mathcal{L}}_{\rm Yukawa} = \sum_{f=u,d,l} \left(\frac{m_f}{v} \kappa_f^h \bar{f} f h + 
\frac{m_f}{v}\kappa_f^H \bar{f} f H 
- i \frac{m_f}{v} \kappa_f^A \bar{f} \gamma_5 f A \right) + \nonumber \\
\left(\frac{V_{ud}}{\sqrt{2} v} \bar{u} (m_u \kappa_u^A P_L +
m_d \kappa_d^A P_R) d H^+ + \frac{ m_l \kappa_l^A}{\sqrt{2} v} \bar{\nu}_L l_R H^+ + \text{h.c.} \right),
\label{Yukawa-1}
\end{eqnarray}
where $\kappa_f^S$ ($S=h,H$ and $A$) are the Yukawa couplings of the fermion $f$ in the 2HDM, which are illustrated 
in Tab.~\ref{yuk_coupl} for the Type-X under consideration. Here, $V_{ud}$ refers to a Cabibbo-Kobayashi-Maskawa 
(CKM) matrix element and $P_{L,R}$ denotes the left- and right-handed projection operators, respectively. The couplings of the two CP-even states $h$ and $H$ to the gauge bosons $VV$ ($V = W^\pm, Z$) are respectively proportional to $\sin(\beta-\alpha)$ and $\cos(\beta-\alpha)$. If we assume that either $h$ or $H$ can be the observed SM-like Higgs boson, the coupling to gauge bosons is obtained for $h$ when $\cos(\beta-\alpha) \rightarrow 0$ and for $H$ when $\sin(\beta-\alpha) \rightarrow 0$. 
Therefore, each scenario can explain the 125 GeV Higgs signal at the LHC. 
Following our previous works \cite{Arhrib:2021xmc,Wang:2021pxc,Arhrib:2021yqf,Li:2023btx}, we shall focus in the present paper on the scenario where $H$ mimics the observed signal with mass $\sim\,125$ GeV (as previously intimated).

\begin{table}[H]
	\centering
	\renewcommand{\arraystretch}{1.2} %
	\setlength{\tabcolsep}{1.2pt}
	\begin{tabular}{|c|c|c|c|} 
    \hline
		& $\kappa_u^{S}$ &  $\kappa_d^{S}$ &  $\kappa_\ell^{S}$  \\   \hline
		$h$~ 
		& ~ $  \cos\alpha/ \sin\beta$~
		& ~ $  \cos\alpha/ \sin\beta$~
		& ~ $  -\sin\alpha/ \cos\beta $~ \\ \hline
		$H$~
		& ~ $  \sin\alpha/ \sin\beta$~
		& ~ $  \sin\alpha/ \sin\beta$~
		& ~ $  \cos\alpha/ \cos\beta$~ \\ \hline
		$A$~  
		& ~ $  \cot \beta $~  
		& ~ $  -\cot \beta $~  
		& ~ $  \tanb $~  \\ \hline
	\end{tabular}
	\caption{Yukawa couplings of the fermions $f=u,d$ and $\ell$ to the neutral Higgs bosons $S=h,H$ and $A$ in the 2HDM Type-X.}
	\label{yuk_coupl}	
\end{table}	

\section{Bounds and constraints in the parameter space scans }\label{psc}
To determine the regions that meet both theoretical requirements and experimental observations, the subsequent numerical explorations of the potential parameter spaces have been carried out: 
\begin{eqnarray}
&~& \mh \in [90, 125]~\gev,\quad \mH = 125~\gev,\quad    \mA \in [90, 125]~\gev \nonumber\\
&~& \mcH \in [90, 125]~\gev,\quad  \sinbma \in [-0.25, 0.05],\quad   \tanb \in [20, 200] \nonumber\\
&~&  m_{12}^2 = \mcH^2\frac{\tanb}{1+\tan^2\beta}~\text{GeV}^2.
\end{eqnarray}

The program \texttt{2HDMC-1.8.0}~\cite{Eriksson:2009ws} was employed to calculate the Higgs Branching Ratios (BRs) and to examine the following theoretical and experimental constraints.
\begin{itemize}
	\item Vacuum stability~\cite{Deshpande:1977rw}:  the scalar potential should be bounded by the conditions $\lambda_1 >0$, $\lambda_2>0$, $\lambda_3>-\sqrt{\lambda_1\lambda_2}$ and $\lambda_3+\lambda_4-\lambda_5>-\sqrt{\lambda_1\lambda_2}$.
	\item Perturbativity constraints~\cite{Kanemura:1993hm, Akeroyd:2000wc}, which imply the condition $|\lambda_{i}|\textless 8\pi$, which holds for $i=1$ to 5. 
    \item Tree-level perturbative unitarity~\cite{Kanemura:1993hm, Arhrib:2000is, Akeroyd:2000wc}: requiring all $2\to2$ scattering processes involving Higgs and gauge bosons at high energy to be unitary.
    \item Oblique EW parameters ($S$, $T$ and $U$)~\cite{Peskin:1990zt, Peskin:1991sw}, which govern the mass splitting between the Higgs states, with the following measured values~\cite{ParticleDataGroup:2022pth}:
\begin{align}
S= -0.02\pm 0.10,\quad T = 0.03\pm 0.12.
\end{align}
The correlation factor between $S$ and $T$ is set to $0.92$ for consistency at $95\%$ Confidence Level (CL). 
\end{itemize}

To account for potential additional Higgs bosons, exclusion bounds are enforced using the \texttt{HiggsBounds-5.9.0} program \cite{Bechtle:2020pkv}, which systematically checks each parameter point against the $95\%$ CL exclusion limits derived from Higgs boson searches conducted by LEP, Tevatron and LHC experiments. 

To ensure agreement with the measurements of the SM-like Higgs state, constraints are enforced using the \texttt{HiggsSignals-2.6.0} program \cite{Bechtle:2020uwn}, which incorporates the combined measurements of the SM-like Higgs boson from LHC Run-1 and Run-2 data. 

In addition, we have cross-validated the above scan results using the latest HiggsTools framework \cite{Bahl:2022uwn}, which integrates updated exclusion limits and signal-strength likelihoods from LHC 13 TeV data within a unified interface.

The constraints of flavour physics are incorporated using  \texttt{SuperIso v4.1} \cite{Mahmoudi:2008tp} using the following observables: 
	\begin{itemize}
		\item BR$(B \to X_s \gamma) = (3.32 \pm 0.15) \times 10^{-4}$ \cite{HFLAV:2022esi},
  		\item	${\rm BR}(B^0\to \mu^+\mu^-)_{\text{(LHCb)}}$=$\left(1.2^{+0.8}_{-0.7}\right)\times 10^{-10}$~\cite{LHCb:2021awg,LHCb:2021vsc}, 
			\item ${\rm BR}(B^0\to \mu^+\mu^-)_{\text{(CMS)}}$=$\left(0.37^{+0.75}_{-0.67}\right)\times 10^{-10}$~\cite{CMS:2022mgd},
           \item	${\rm BR}(B_s\to \mu^+\mu^-)_{\text{(LHCb)}}$ = $\left(3.09^{+0.46}_{-0.43}\right)\times 10^{-9}$~\cite{LHCb:2021awg,LHCb:2021vsc},
			\item	${\rm BR}(B_s\to \mu^+\mu^-)_{\text{ (CMS)}}$=$\left(3.83^{+0.38}_{-0.36}\right)\times 10^{-9}$~\cite{CMS:2022mgd},
   \item BR$(B \to \tau \nu) = (1.09 \pm 0.20) \times 10^{-4}$ \cite{HFLAV:2022esi}. 
	\end{itemize}

\begin{figure}[h!]
	\hspace*{-0.3cm}
	\includegraphics[scale=0.5]{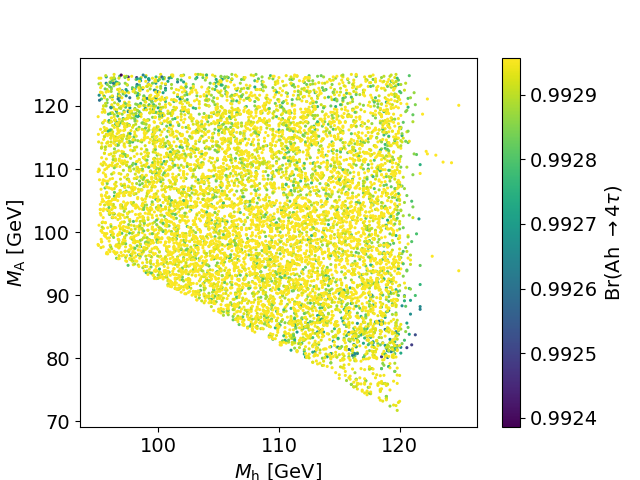}	
 \hspace*{-0.4cm}
	\includegraphics[scale=0.5]{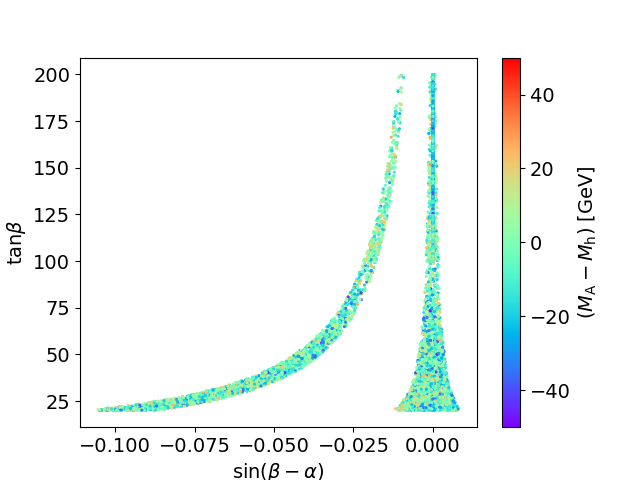}
	\caption{Scatter plot of ${\rm{BR}}(h \to \tau^+\tau^-)\times {\rm{BR}}(A \to \tau^+\tau^-)$ as a function of $\mh$ and $\mA$ (left panel) and  $\mA - \mh$ as a function of  $\sinbma$ and $\tanb$ (right panel).}
	\label{fig:scan}
\end{figure}

In the left panel of Fig.~\ref{fig:scan}, we show the overall BR of the light Higgs $h$ and the pseudoscalar $A$ decaying into pairs of $\tau$ leptons, i.e., BR$(Ah \to 4\tau)$, within the allowed $(\mh, \mA$) plane of the 2HDM Type-X. Such a significant BR is attributable to the enhancement of the $\hboson/\Aboson$ couplings to leptons at large $\tanb$.  The parameter spaces for $\sinbma$ and $\tanb$  are shown in the right panel, where the colour bar is $\mA-\mh$. It can be seen that most of the points are within the $\mA < \mh$ region.

\begin{figure}[h!]
	\hspace*{-0.3cm}
	\includegraphics[scale=0.5]{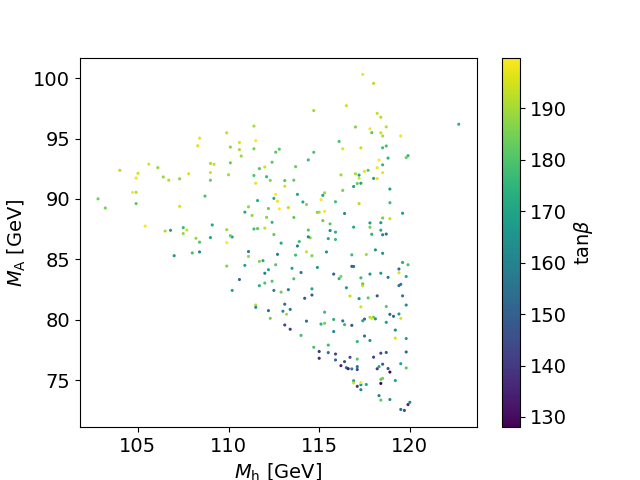}	
	\caption{Scatter plot of $\tanb$ as a function of $\mh$ and $\mA$ when the $a_\mu$ effect is considered in addition to the constraints imposed in the previous figure.}
	\label{fig:g-2}
\end{figure}

When the constraint on $a_\mu$ is considered, there are two types of contributions to $\Delta a_{\mu}$, which are one-loop corrections from $H$, $A$ and $H^{\pm}$~\cite{Chun:2016} as well as two-loop Barr-Zee contributions with heavy fermions~\cite{Ilisie:2015}.
Only in the large $\tanb$ region is $\Delta a_{\mu}$ large enough to explain the latest FNAL measurements. The surviving parameter space is shown in Fig.~\ref{fig:g-2}. Compared with Fig.~\ref{fig:scan} (left), most of the parameter space has been excluded: in fact, only points with $\tanb>130$ are available and the accessible ($\mh, \mA$) regions are rather sparse.

To test the allowed parts of the parameter space, nine BPs are given in Tab.~\ref{t:BPs}, in terms of the 2HDM Type-X parameters entering our production and decay process, including the corresponding $pp \to Z^{*} \to hA \to 4\tau$ cross sections at the LHC with $\sqrt{s} =14$ TeV. Based on this table, we select two scenarios. For BPs~1--4, where $\tanb$ is always below 100 and $\sin(\beta-\alpha)$ is close to but not exactly zero, we note that $a_\mu$ remains a concern (till new measurements are performed). For BPs~5--9, the model is in the alignment-limit scenario, where $\sin(\beta-\alpha) = 0$ and $\tanb$ is very large while $a_\mu$ can be explained well (with the present measurements).

\begin{table}[h!]
	\begin{center}
			\begin{tabular}{| c| c| c| c| c| c| c|}
				\hline
				~ & $\mh$ & $\mA$ & $\mcH$ & $\tanb$ & $\sin(\beta - \alpha)$ &  $\sigma_{4\tau,~14~{\rm TeV}}$ (fb) \\
				\hline  
BP1 & 95.10 & 121.70 & 95.02 & 24.19 & $-0.076$ & 38.76 \\
\hline
BP2 & 120.10 & 118.50 & 119.80 & 41.25 & $-0.003$ & 27.89 \\
\hline
BP3 & 99.52 & 102.30 & 99.53 & 60.71 & $-0.034$ & 52.01 \\
\hline
BP4 & 109.10 & 89.07 & 109.00 & 84.43 & $-0.025$ & 55.23 \\
\hline
BP5 & 105.70 & 117.40 & 105.60 & 105.30 & $-0.001$ & 35.62 \\
\hline
BP6 & 113.00 & 103.20 & 113.00 & 137.80 & 0.0 & 39.79 \\
\hline
BP7 & 109.70 & 115.30 & 109.70 & 143.00 & 0.0 & 34.24 \\
\hline
BP8 & 107.00 & 85.31 & 107.00 & 166.10 & 0.0 & 60.71 \\
\hline
BP9 & 116.20 & 111.20 & 116.20 & 196.70 & 0.0 & 31.79 \\
   				\hline
			\end{tabular}
			\caption{2HDM Type-X input parameters and Leading Order (LO) cross sections  (at the parton level with $\sqrt{s} =14$ TeV) are presented for each BP. The unit of all masses is GeV and we fix the SM-like Higgs boson mass as $\mH$ = 125 GeV.
              (Also recall that $m_{12}$ has been taken as a derived quantity, function of $\mcH$ and $\tanb$.) }\label{t:BPs}
	\end{center}
\end{table}

\section{Collider phenomenology}
In this section, we present a detailed MC analysis for both signal and background events at the detector level. 
In the $pp\to Z^{*}\to hA\to \tau^{+}\tau^{-}\tau^{+}\tau^{-}\to l\nu_l l\nu_l \tau_h \tau_h$ process, i.e., the light Higgs (pseudo)scalar decays into two $\tau$ leptons and one of the $\tau$'s decays into a charged lepton and neutrino, while the other $\tau$ decays hadronically (thus labelled as $\tau_h$). 

In the CMS experiment, there are two search strategies for light scalars in term of topologies of final states and different topologies have different SM backgrounds. The first strategy targets a light Higgs boson h (or a pseudoscalar A) with a mass in the range of 5-15 GeV, produced in pairs from the decay process $H \to hh$~\cite{CMS:2015twz, CMS:2017dmg, CMS:2019spf}. Subsequently, each light Higgs boson decays into two tau leptons, with at least one tau lepton decaying into a muon. Since the light scalars from Higgs decay are highly boosted, the 4 tau final states might form a topology with a pair of two collimated tau decay products. Thus, the backgrounds should include Drell--Yan di-muon pairs associated with jets, $W$+jets, WW, WZ, ZZ, and QCD multi-jets. In the second strategy, the light scalar bosons are assumed to be in the mass range of 15--62.5~GeV~\cite{CMS:2017dmg, CMS:2018ams} and the final states with $2\mu 2\tau$, $2\mu 2b$ and $4\mu$ are analyzed. The four objects in the final states are well separated. For these final states of signals, the irreducible $ZZ$ background, and reducible $Z$+jets, $t\bar{t}$ and $WZ$+jets backgrounds~\cite{CMS:2017dmg} should be taken into account. In this work, the light Higgs boson and the pseudoscalar are around 100~GeV and they are not highly boosted from their production processes and the majority of their decay products are well-separated. Therefore, we primarily adopt the analysis of the second strategy. Thus, we will consider the following SM backgrounds:
$t \overline{t}\to l \nu_l l \nu_l b \overline{b}$, 
$W^\pm t b\to l \nu_l l \nu_l b \overline{b}$ (where the $W^\pm$ boson and the $b$ jet do not originate from a top quark resonance), 
$W W j j\to l \nu_l l \nu_l j j$, 
$Z j j\to l \nu_l l \nu_l j j$, 
$Z Z\to l \nu_l l \nu_l \tau_h \tau_h$, 
$t \overline{t} Z\to l \nu_l l \nu_l b \overline{b} \tau \tau$, 
$t \overline{t} Z Z\to l \nu_l l \nu_l l \nu_l b \overline{b} \tau \tau$, and 
$t \overline{t} W W\to l \nu_l l \nu_l l \nu_l j j$.

To suppress the huge SM background process $Z j j$, we deliberately choose two same-sign (SS) leptons plus two hadronic $\tau$'s (for signal events, this is about $10\%$ of the total number of $4\tau$ ones). As a result, the final state consists of two SS leptons and two (light) jets.

In the experimental analysis of same sign lepton final states, for example  \cite{CMS:2018agk}, the background typically include the misidentified leptons from jets or photons, and the sign mismeasured leptons. For the misidentified leptons, due to challenging difficulties, precise information can only be estimated by using control samples of collision data. In this analysis, we can not include such typs of backgrounds. While for the sign mismeasured leptons, the rate of sign mismeasurement depends upon the region of detector, and it was found to be around $1.5\times 10^{-5}$ in the inner ECAL barrel region for instance. Thus, the contamination of some background events like $Z jj$ can be estimated, although it has a huge cross section. Considering the fact that when a realistic work point of hadronic tau tagging is 0.6 or so \cite{ATLAS:2022aip}, the rejection factor of faked tau from jets can reach to 50 or above, there is no doubt that the background from sign mismeasured leptons is well under control.

\subsection{Event generation and selection}
We use \texttt{MadGraph5$_{-}$aMC@NLO v3.4.0}~\cite{Alwall:2014hca} to calculate cross sections and generate both signal and background events at the parton level. The following acceptance cuts are applied for signal and background generation:
\begin{eqnarray}\label{parton_cuts}
|\eta (l,j)|< 2.5,~p_{\text{T}}(l,j)> 10~\gev,~E_T^{\text{miss}}>10~\text{GeV},~\Delta R(ll,lj,jj)>0.4.   
\end{eqnarray}

The parton-level events are then passed to Pythia~\cite{Sjostrand:2006za,Sjostrand:2014zea} to simulate initial- and final-state radiation (i.e., QED and QCD emissions), parton showering, hadronisation and heavy flavour decays. We further use \texttt{Delphes v3.4.2}~\cite{deFavereau:2013fsa} to simulate the detector effects. For each event, the anti-$k_t$ jet algorithm~\cite{Cacciari:2008gp}
is used to cluster the jets with the jet parameter $\Delta R = 0.4$ in the \texttt{FastJet} package~\cite{Cacciari:2011ma}. The following kinematic cuts are further applied in order to emulate the detector acceptance:
\begin{eqnarray}\label{detector_cuts}
|\eta (l,j)|< 2.5,~p_T(l,j)> 10~\gev.   
\end{eqnarray}

The leptons are used for triggering purposes, so their kinematics must satisfy the trigger requirements\footnote{The CMS collaboration has developed a double-muon trigger to search for low $P_T$ muons from $B$ meson decays that are close together. A similar approach has been adopted recently to search for electron pairs from $B$ meson decays, by tightening the topological selection criteria on the two Level-1 electron objects. In our recent paper~\cite{Arhrib:2023apw}, we explored the possibility of also developing an electron-muon trigger to search for low $P_T$ leptons ($P_T \sim 10$ GeV).}.

Next, we select a final state consisting of two leptons and two hadronic tau jets from detector-simulated events. The final cut table shows that the dominant background is the irreducible $ZZ$ process, whose tau jet distribution closely resembles that of the signal. In the analysis below, in order to have a close look at the effects of kinematic observables, we leave the discussion of effects of hadronic tau tagging in the discussion section. 

The signal results are presented in Tab.~\ref{t:CS_signal}, and the relevant background processes are summarised in Tab.~\ref{t:bg}. The number of background events is approximately $10^5$ times larger than that of the signal events, with the dominant background processes at this stage being $pp\to t {\bar t}$ and $pp \to Z jj$. 

To further suppress the huge SM backgrounds, we impose the selection condition requiring SS leptons. After applying this criterion, the number of signal and background events becomes comparable, with the dominant background contributions arising from the $pp\to t {\bar t}$ and $pp \to ZZ$ processes.

The cross sections of the signal and background processes after the detector acceptance cuts, the selection condition assuming a $2l2j$ final state, along with the two SS leptons requirement are shown in Tabs.~\ref{t:CS_signal} and~\ref{t:bg}, respectively.

\begin{table}[H]
	\begin{center}
			\begin{tabular}{| c| c| c| c| c| c| c| c| c| c|}
				\hline
			$\sigma$ (fb) & BP1  & BP2   & BP3   & BP4  & BP5  & BP6  & BP7  & BP8  & BP9  \\
				\hline  
     Parton level & 10.96 & 8.75 & 13.66  & 14.16 & 10.43 & 11.24 & 10.09 & 14.79 & 9.49\\
                    \hline
           Selection of $2l2j$  & 3.02 &  2.59 & 3.66 & 3.68  & 2.87 & 3.20 &2.95 & 3.83 & 2.67\\
                    \hline
           Selection of SS $2l$ & 1.196 &  1.494 & 1.522 & 1.498 & 1.007 &1.239 &1.141 &1.301 &1.295\\
				\hline  
			\end{tabular}
			\caption{Cross sections for all signal BPs at the parton level as well as the detector level after selection cuts at the LHC with $\cmsfourteen$.
            }\label{t:CS_signal}
	\end{center}
\end{table}

\begin{table}[H]
	\begin{center}
			\begin{tabular}{|c| c| c| c| c| c| c| c| c|}
				\hline
				$\sigma$ (fb) & $t\bar{t}$ & $W^\pm tb$ & $W^+W^-jj$ & $Zjj$ & $ZZ$ & $t\bar{t}Z$ & $t\bar{t}ZZ$ & $t\bar{t}W^+W^-$ \\
				\hline
	    Parton level &  16060 & 518.3 & 1053 & 317600 & 18.89 & 0.49 & $1.14\times10^{-4}$ & 0.02\\
				\hline
    	Selection of $2l2j$      &  8787.7 & 289.9 & 530.1 & 151086 & 10.0 & 0.33 & $1.1\times10^{-4}$ & 0.018\\
				\hline  
    	  Selection of SS $2l$     &  19.43 & 0.62 & 1.99 & 0 & 2.51 & 0.079 & $3.3\times10^{-5}$ & $7.6\times10^{-3}$\\
				\hline  
			\end{tabular}
			\caption{Cross sections for all backgrounds at the parton level as well as the detector level after selection cuts at the LHC with $\cmsfourteen$.
            }\label{t:bg}
	\end{center}
\end{table} 

\begin{figure}[ht]
\begin{minipage}{0.47\textwidth}
      \begin{center} 
         \includegraphics[height=5.0cm]{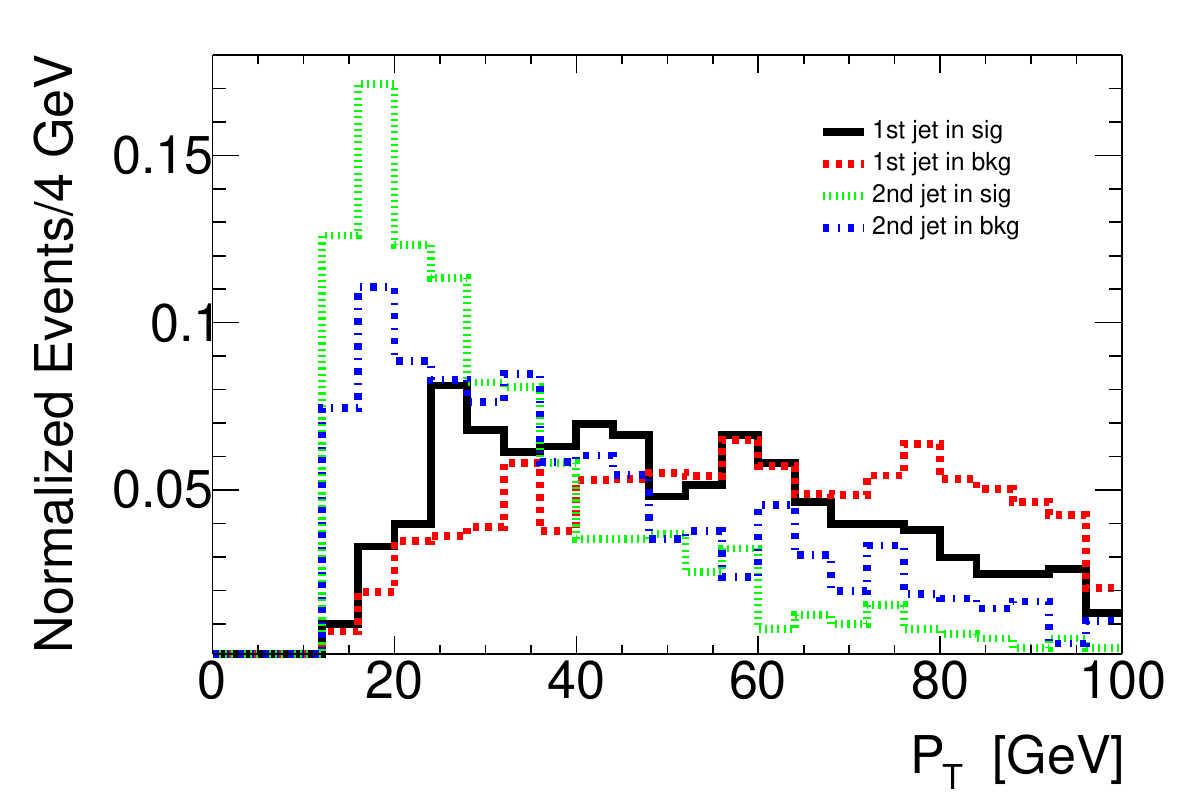}	
                              (a)	\\
      \end{center}
\end{minipage}
\begin{minipage}{0.47\textwidth}
      \begin{center} 
         \includegraphics[height=5.0cm]{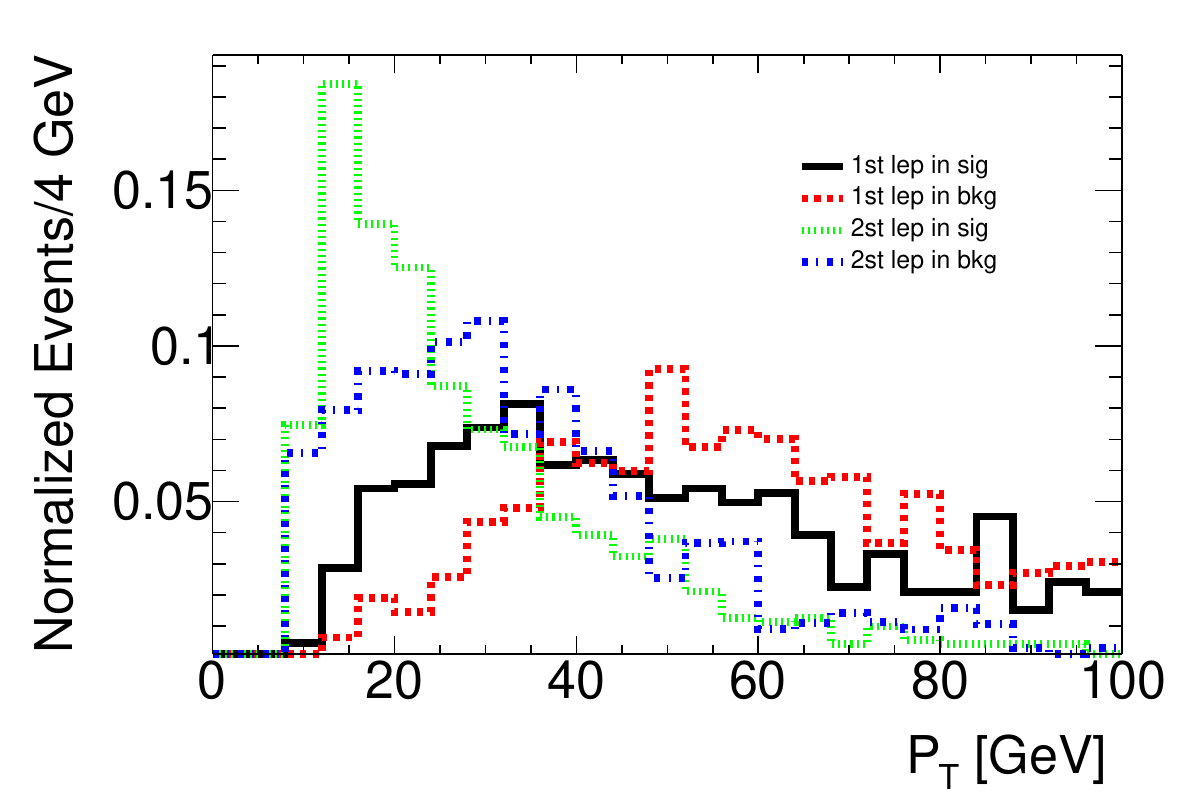}	
                              (b)	\\
      \end{center}
\end{minipage}

 \caption{Normalised distributions in transverse momentum of the first- and second-leading jets (a)  and leptons (b) for BP4 and backgrounds at the LHC with $\cmsfourteen$   and $L~=~\threefbm$.}\label{f:pt}
\end{figure}

In Fig.~\ref{f:pt}, the transverse momentum distributions of the (second-)leading jets (frame (a)) and leptons (frame (b)) are presented for the signal (e.g., BP4) and all background events. It is clear that the transverse momenta of the signal final states are always much softer than those of background events. This is because they emerge from rather light Higgs bosons, whose energies are capped at masses of the order of 100 GeV.

\subsection{Event reconstruction}
To further improve the signal-to-background discrimination, we reconstruct some kinematic features of signal events. However, it is extremely difficult to completely reconstruct the momentum of each light Higgs state. This is because each Higgs boson decays into two $\tau$'s, with one of the $\tau$'s decaying to a lepton and two neutrinos, plus two $\tau$ neutrinos emerging from hadronic $\tau$ decays, leading to the presence of six neutrinos in a signal event.

\begin{figure}[t!]
	\centering
	\begin{minipage}{8cm}
		\centering
		\includegraphics[width=8cm]{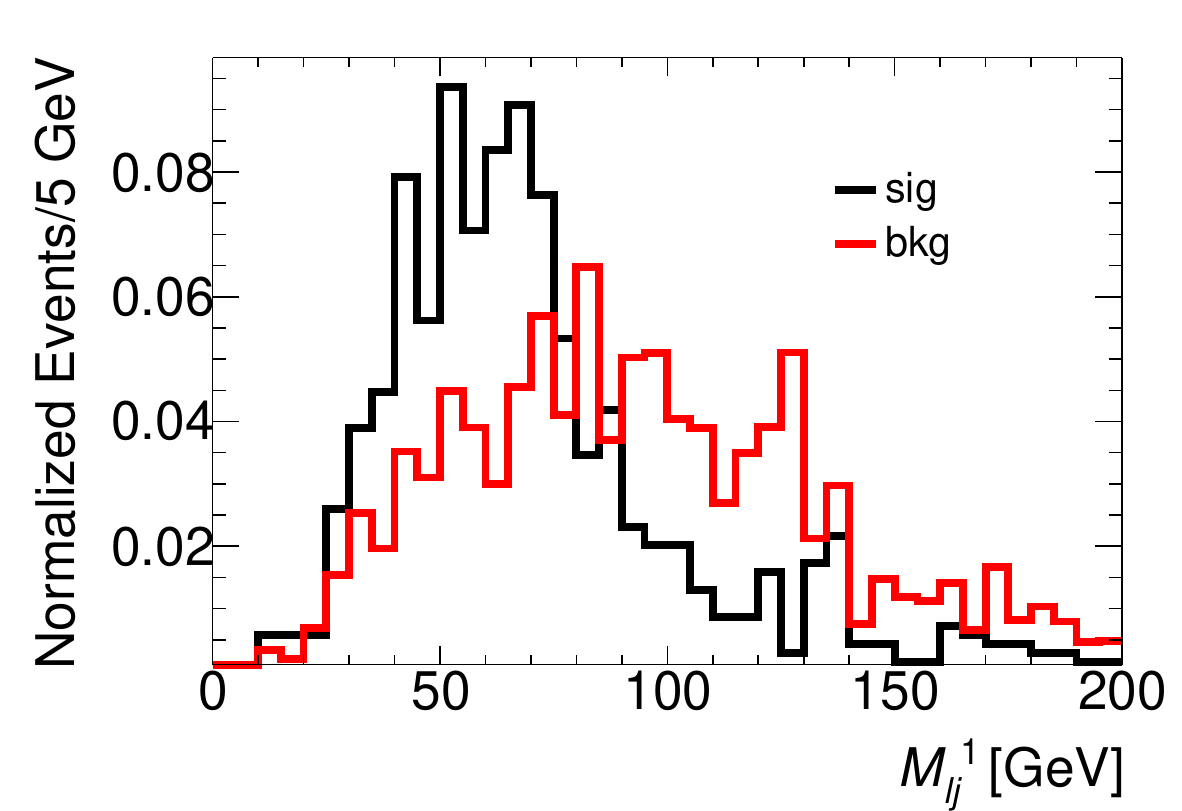}
        (a)
	\end{minipage}
    \begin{minipage}{8cm}
		\centering
		\includegraphics[width=8cm]{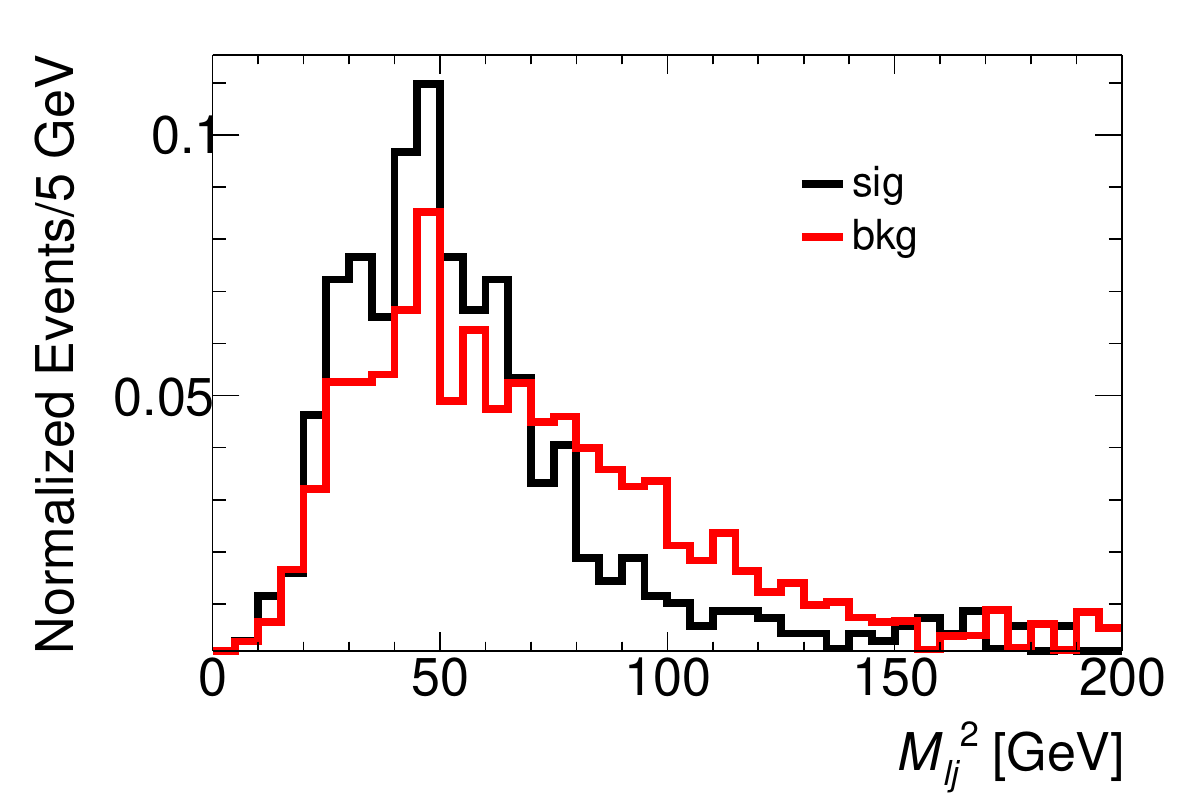}
        (b)
	\end{minipage}
    \begin{minipage}{14cm}
		\centering
		\includegraphics[width=14cm]{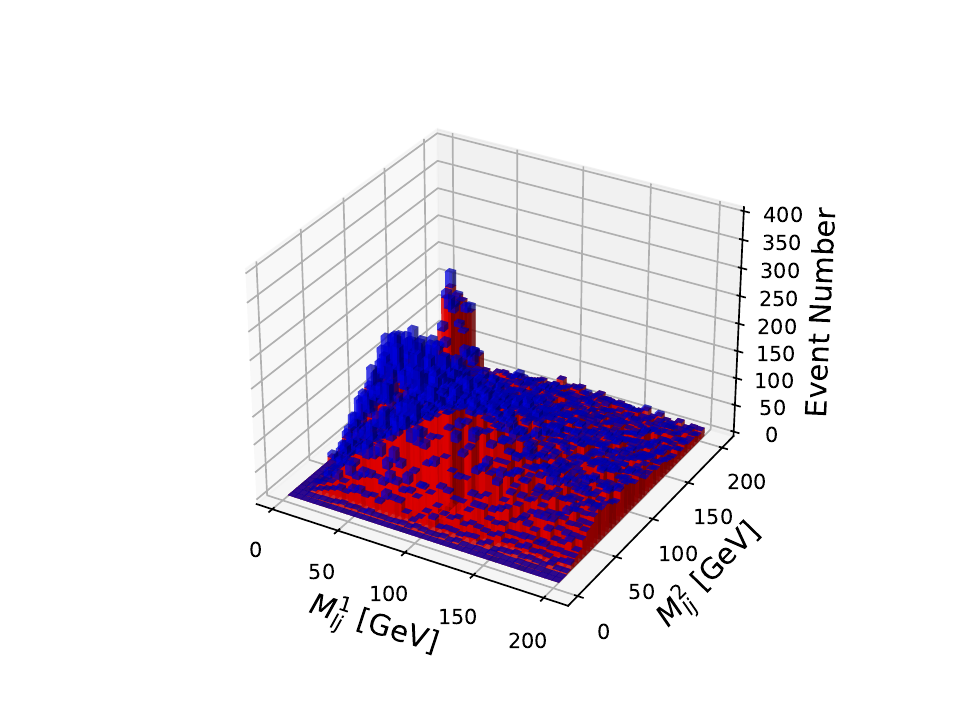}
                (c)
	\end{minipage}
\caption{Distributions of the invariant mass of a lepton and a jet for BP4 and background processes at the LHC with $\cmsfourteen$ and an integrated luminosity of $L = \threefbm$. The reconstructed masses, obtained from the best $\chi^2$ fit, are denoted by $M_{lj}^1$ and $M_{lj}^2$, and are shown in normalized form in panels (a) and (b), respectively. Panel (c) displays the unnormalized distributions (corresponding to $L = \threefbm$) with the correlation between $M_{lj}^1$ and $M_{lj}^2$.  In the bottom frame, the signal (background) is in blue(red).}\label{f:Mh1Mh2}
\end{figure}

Nevertheless, since the lepton momenta and the missing $E_T$
(MET), where $E_T$ refers to transverse energy, are all very small, we can approximately reconstruct the light Higgs bosons with only one lepton and one jet. Thus, there are two possible combinations with two leptons and two $\tau$ jets in each event. By defining a suitable $\chi^2$,
\begin{equation}\label{eq:h1h2}
\chi^2 = \frac{{(M_{{\text{lj}}}^1 - \mh)}^2}{\sigma_h^2} + \frac{{(M_{\text{lj}}^2 - \mA)}^2}{\sigma_A^2},
\end{equation}
where $M_{h(A)}$ are the masses of the $h$($A$) (pseudo)scalars, and $\sigma_{h(A)}$ are the uncertainties in determining the mass peak of $h$($A$). In practice, it is found that when the parameters $\sigma_{h(A)}$ are taken as either the decay widths of h(A) or the experimental uncertainties, the result is almost the same. Therefore, in this analysis, we fix $\sigma_{h(A)}=1$ GeV. We pair the two leptons and two jets to find a combination that minimizes this, and then assign these to $M_{lj}^1$ and $M_{lj}^2$, respectively. Since we did not consider neutrinos during such a mass reconstruction, the final $M_{lj}^{1,2}$ values should be smaller than the real $M_{h/A}$ ones; see Figs.~\ref{f:Mh1Mh2} frames (a) and (b). Taking again BP4 as an example, the reconstructed light Higgs mass distributions are shown in Fig.~\ref{f:Mh1Mh2}. A weak positive correlation is observed between these two variables (with a value of $r$ being 0.27), as exemplified in Fig.~\ref{f:Mh1Mh2} frame (c).

\begin{figure}[t!]
	\centering
	\begin{minipage}{8cm}
		\centering
		\includegraphics[width=8cm]{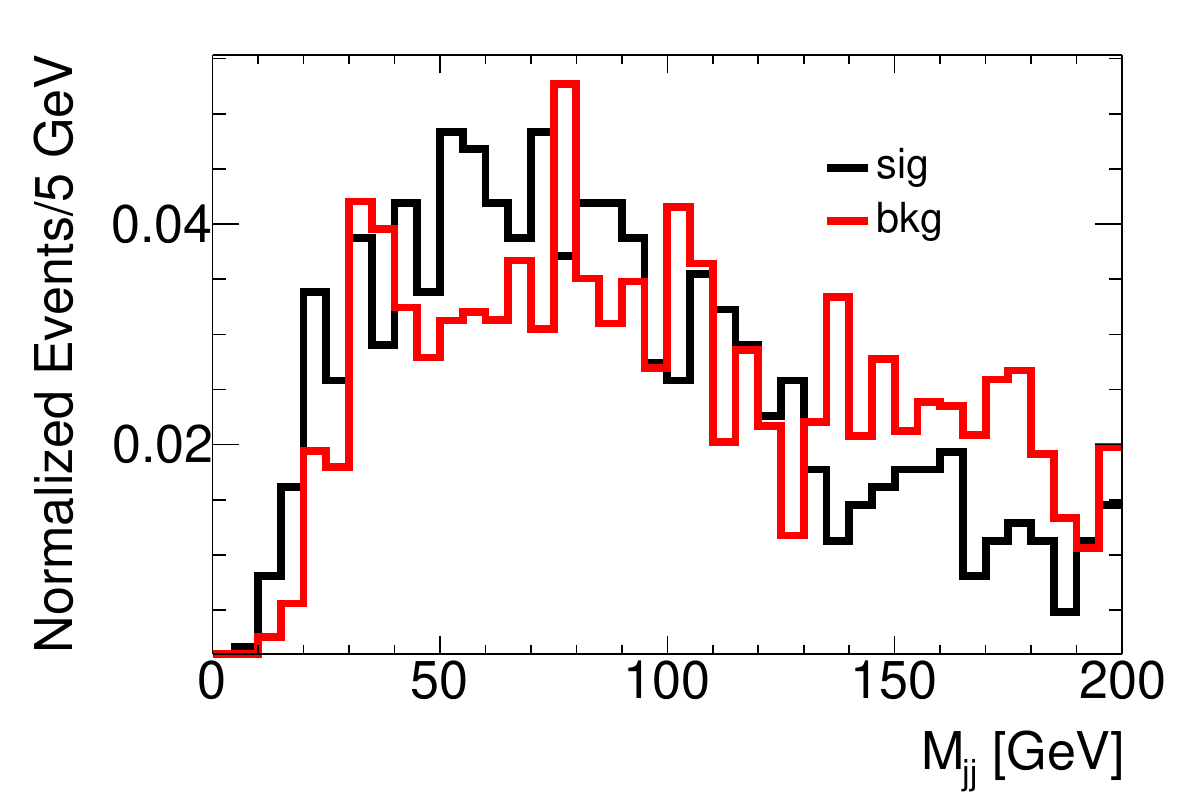}
                             (a)	\\
	\end{minipage}
    \begin{minipage}{8cm}
		\centering
		\includegraphics[width=8cm]{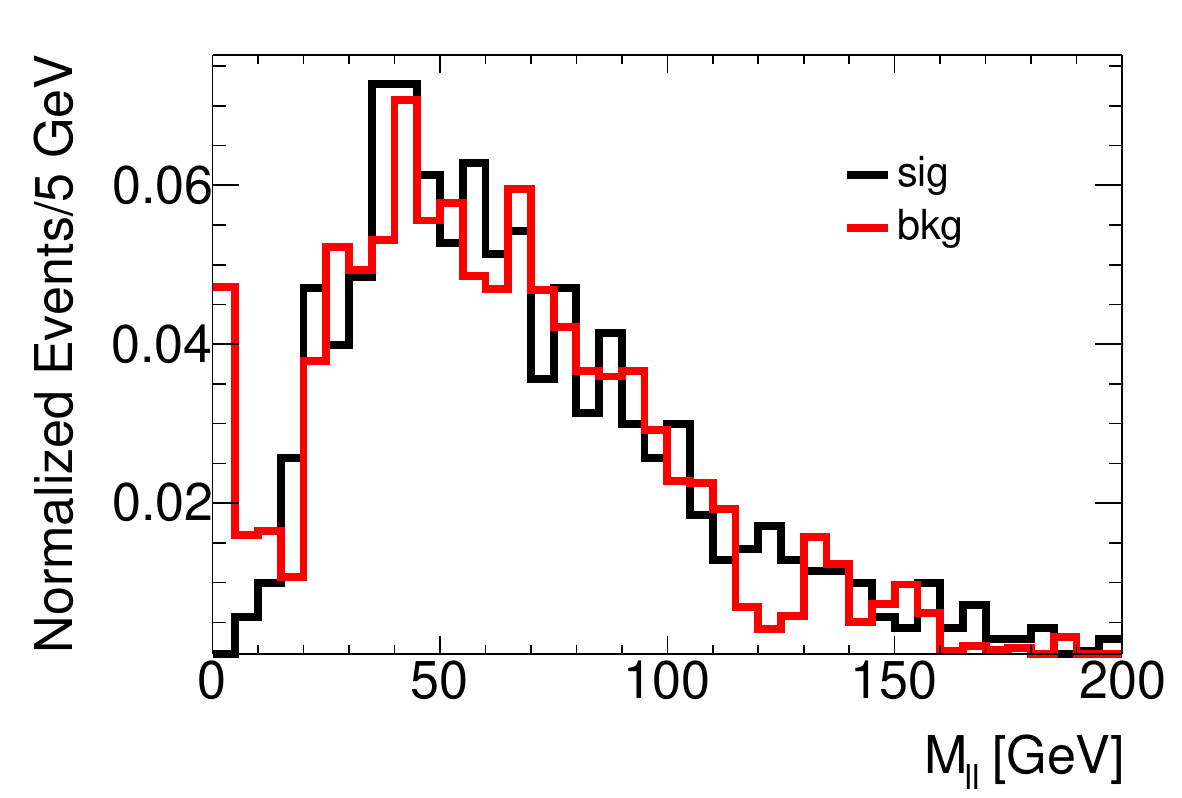}
                             (b)	\\
	\end{minipage}
\caption{Normalised distributions in invariant mass of lepton (a) and jet (b) pairs for BP4 and backgrounds at the LHC with $\cmsfourteen$  and $L~=~\threefbm$.}\label{f:MllMjj}
\end{figure}

In order to further separate the signal from background events, we can also combine the two leptons and two jets, denoting their invariant masses as $M_{ll}$ and $M_{jj}$, respectively. For some SM background processes, these lepton and/or jet pairs always come from some heavy resonance, i.e., a top (anti)quark and a $W^\pm$ or $Z$ boson; the background spectra should have peaks in specific mass regions. However, such a feature is largely removed when selecting SS leptons, as shown in Fig.~\ref{f:MllMjj}, so it cannot really be exploited here.

\begin{figure}[!h]
	\centering
\begin{minipage}{0.47\textwidth}
      \begin{center} 
         \includegraphics[height=5.0cm]{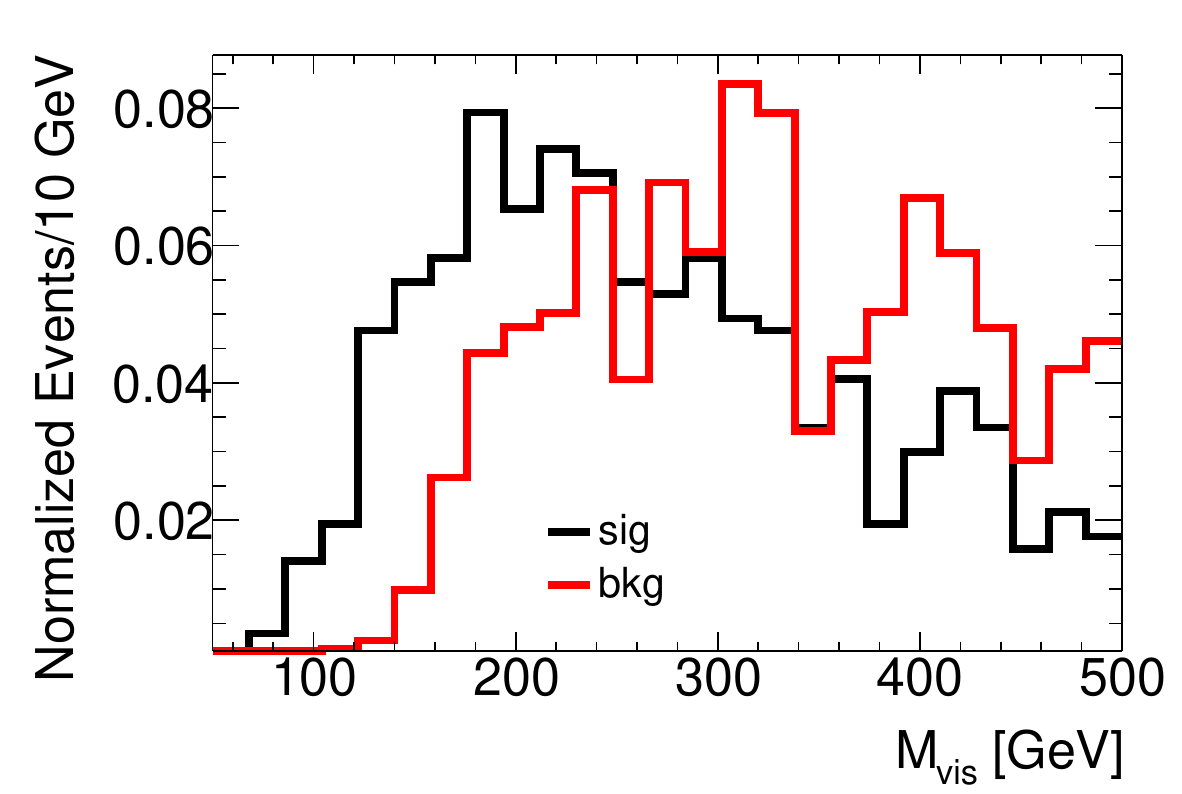}	
                     (a)	\\
      \end{center}
\end{minipage}
\begin{minipage}{0.47\textwidth}
      \begin{center} 
         \includegraphics[height=5.0cm]{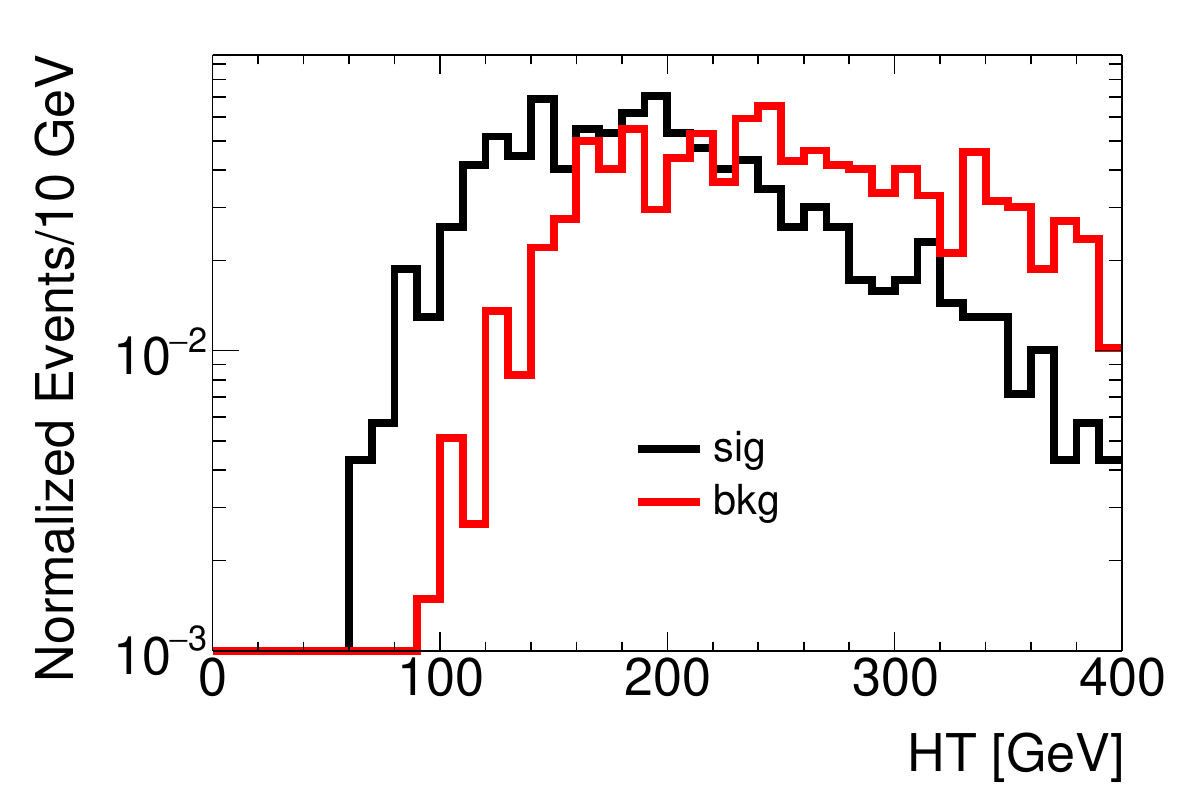}	
                     (b)	\\
      \end{center}
\end{minipage}
\caption{Normalised distributions in visible invariant mass (a) and $H_T$ (b) for BP4 and backgrounds at the LHC with $\cmsfourteen$  and $L~=~\threefbm$.}\label{f:MHHT}
\end{figure}

\begin{figure}[!h]
	\centering
\begin{minipage}{0.47\textwidth}
      \begin{center} 
         \includegraphics[height=5.0cm]{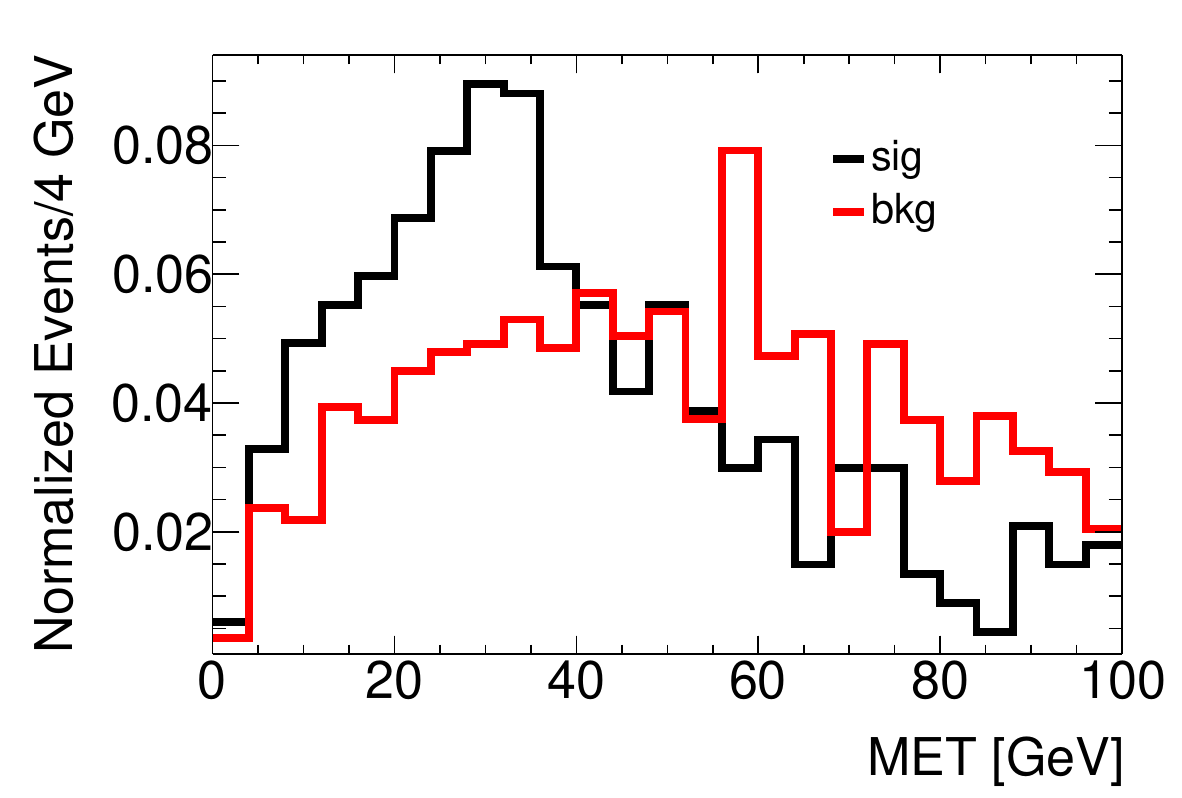}	
                     (a)	\\
      \end{center}
\end{minipage}
\begin{minipage}{0.47\textwidth}
      \begin{center} 
         \includegraphics[height=5.0cm]{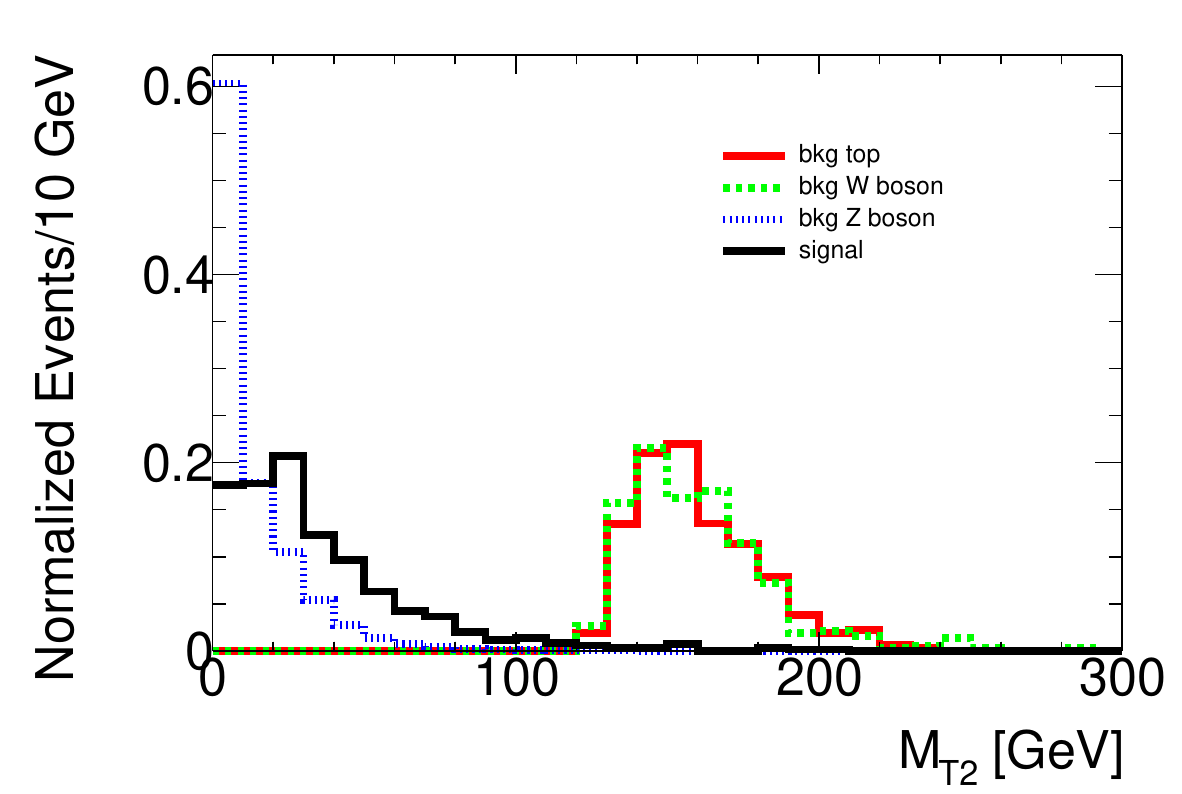}	
                     (b)	\\
      \end{center}
\end{minipage}
\caption{Normalised distributions in missing transverse energy (a) and $\mttwo$ (b) for BP4 and backgrounds  at the LHC with $\cmsfourteen$  and $L~=~\threefbm$. In the right frame we show the three classes of background identified in the text separately.}\label{f:MET}
\end{figure}

We can further reconstruct the visible invariant mass, i.e.\ the one made up of the two leptons and two jets, $M_{\rm vis}$, but without any MET. In the signal, these particles come from a virtual resonance $Z^{*}$, hence $M_{\rm vis}$ should be smaller than in the backgrounds, where this is not normally the case; the latter thus yielding a larger value; see Fig.~\ref{f:MHHT} frame (a). 
Another useful kinematic variable is $H_T$, which is the scalar sum of the transverse momenta of the final-state objects, shown in Fig.~\ref{f:MHHT} frame (b). These two observables are generated in a very similar way between signal and background, which is evident by comparing the two frames in this figure.

In our process, two neutrinos come from leptonic $\tau$ decays, hence are very soft because the $\tau$ leptons are from cascade decays. There are also neutrinos in the backgrounds from $W^\pm$, $Z$ boson or top (anti)quark (semi)leptonic decays, which have momenta higher than those of the neutrinos in signal events. The MET distribution is shown in Fig.~\ref{f:MET} (a). Furthermore, the $\mttwo$ variable is often used to track the masses of unseen pairs of particles~\cite{Lester:1999ma, Barr:2003ma}, which is defined as:
$$\mttwo^{2} \equiv \underset{\slashed{p_{1}}+\slashed{p_{2}}=\slashed{\pt}}{\text{min}}[\text{max}\{\mt^{2}(\pt^{lj^{1}}, \slashed{p_{1}}), \mt^2(\pt^{lj^{2}},\slashed{p_{2}})\}].$$
We thus reconstruct the  $\mttwo$ variable for both signal and background, while splitting the latter into three classes: top (including $t\bar{t}$, $t\bar{t}W^{+}W^{-}$, $t\bar{t}Z$, $t\bar{t}ZZ$ events), $W^\pm$ boson (including $W^{+}W^{-}jj$, $W^{\pm}tb$ events) and $Z$ boson (including $ZZ$ events). The $\mttwo$ distributions are shown in Fig.~\ref{f:MET} (b). It can be seen that those of the signal and $Z$ boson background are below 100 GeV or so, while they are close to 150 GeV or so for the top quark and $W^\pm$ boson processes. Thus, the $\mttwo$ variable provides an efficient discrimination, at least, against the top quark and $W^\pm$ boson backgrounds.

\subsection{Multivariate analysis}

To optimise the signal and background distinction, we exploit a ML analysis, that is, a Gradient-Boosted Decision Tree (GBDT) approach which is further applied, and implemented in the Toolkit for Multivariate Data Analysis (TMVA) with ROOT~\cite{Therhaag:2010zz}. 

\begin{figure}[ht]
\begin{minipage}{0.47\textwidth}
      \begin{center} 
                     (a)	\\
         \includegraphics[height=5.0cm]{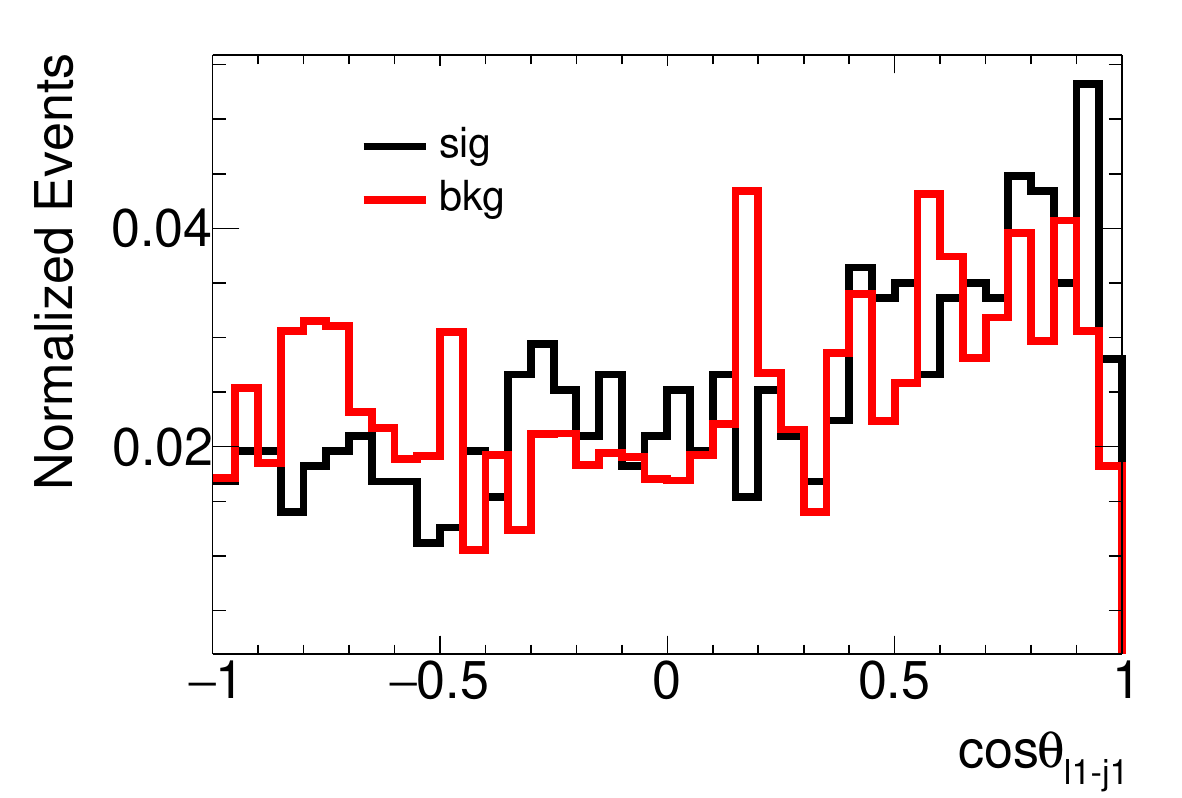}	
      \end{center}
\end{minipage}
\begin{minipage}{0.47\textwidth}
      \begin{center} 
                     (b)	\\
         \includegraphics[height=5.0cm]{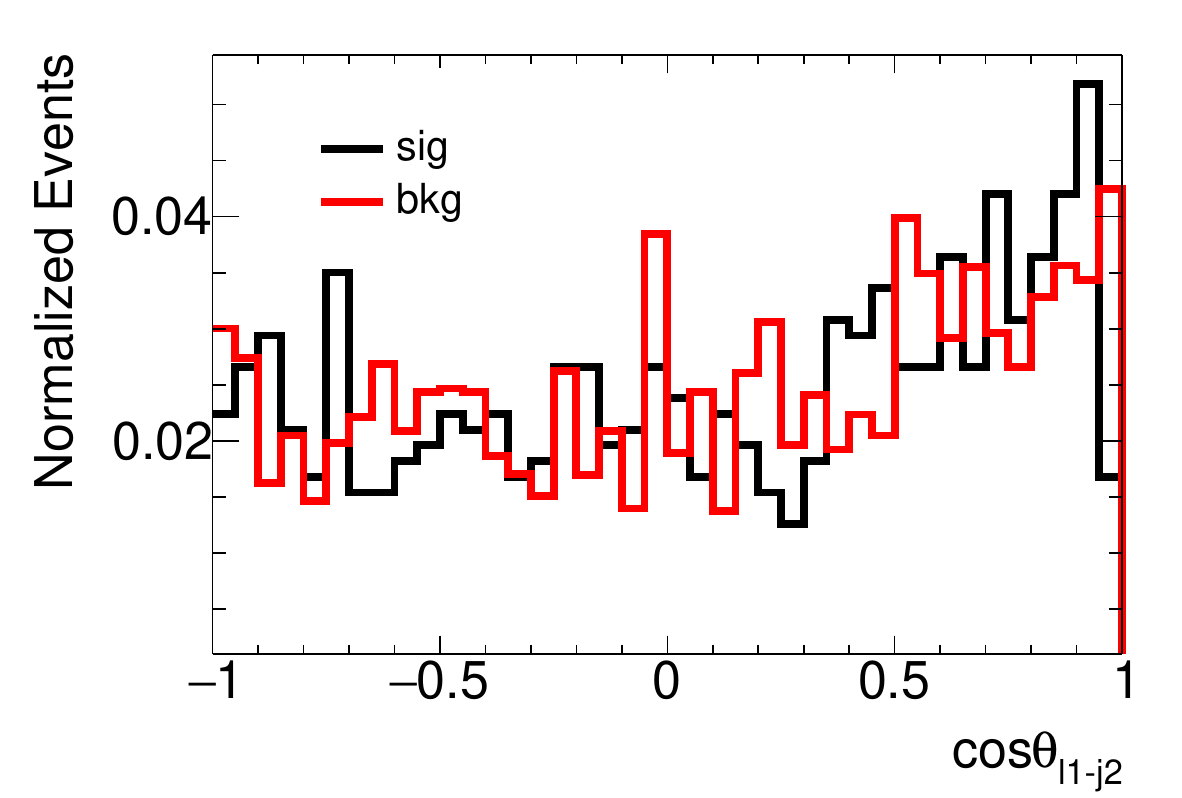}	
      \end{center}
\end{minipage}

\begin{minipage}{0.47\textwidth}
      \begin{center} 
                     (c)	\\
         \includegraphics[height=5.0cm]{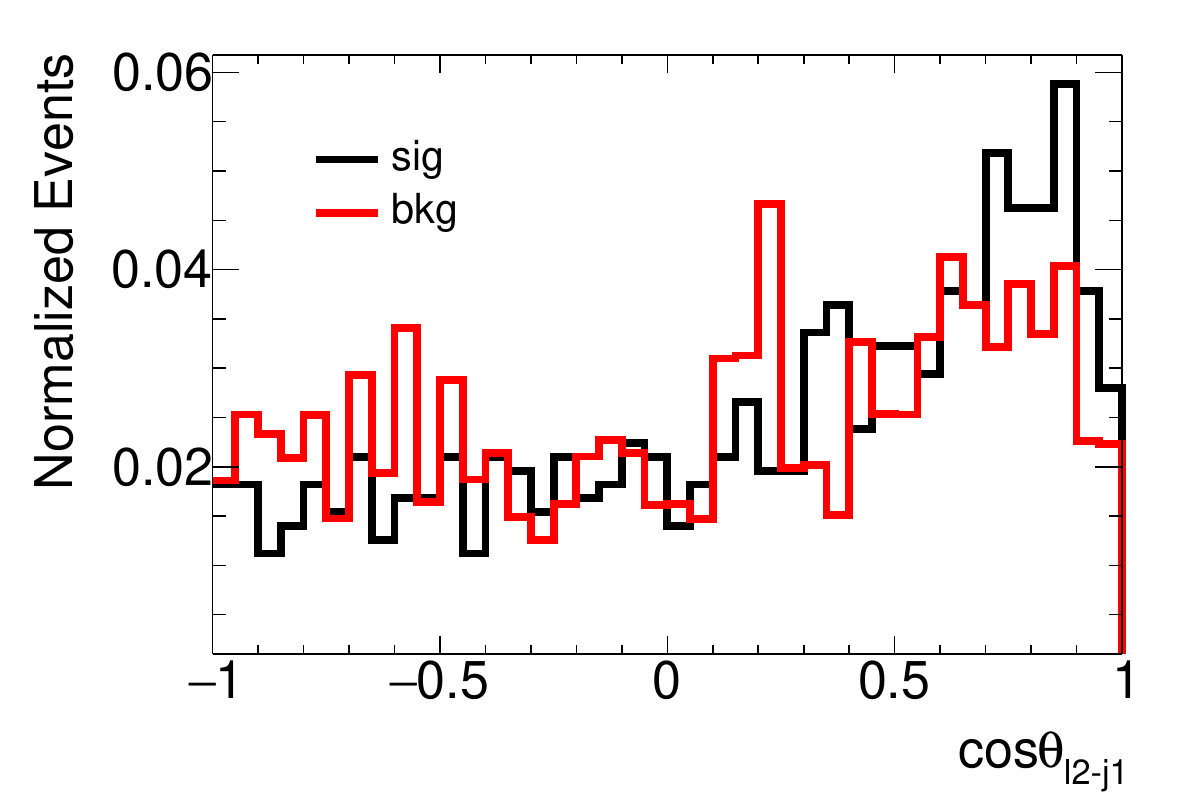}	
      \end{center}
\end{minipage}
\begin{minipage}{0.47\textwidth}
      \begin{center} 
                     (d)	\\
         \includegraphics[height=5.0cm]{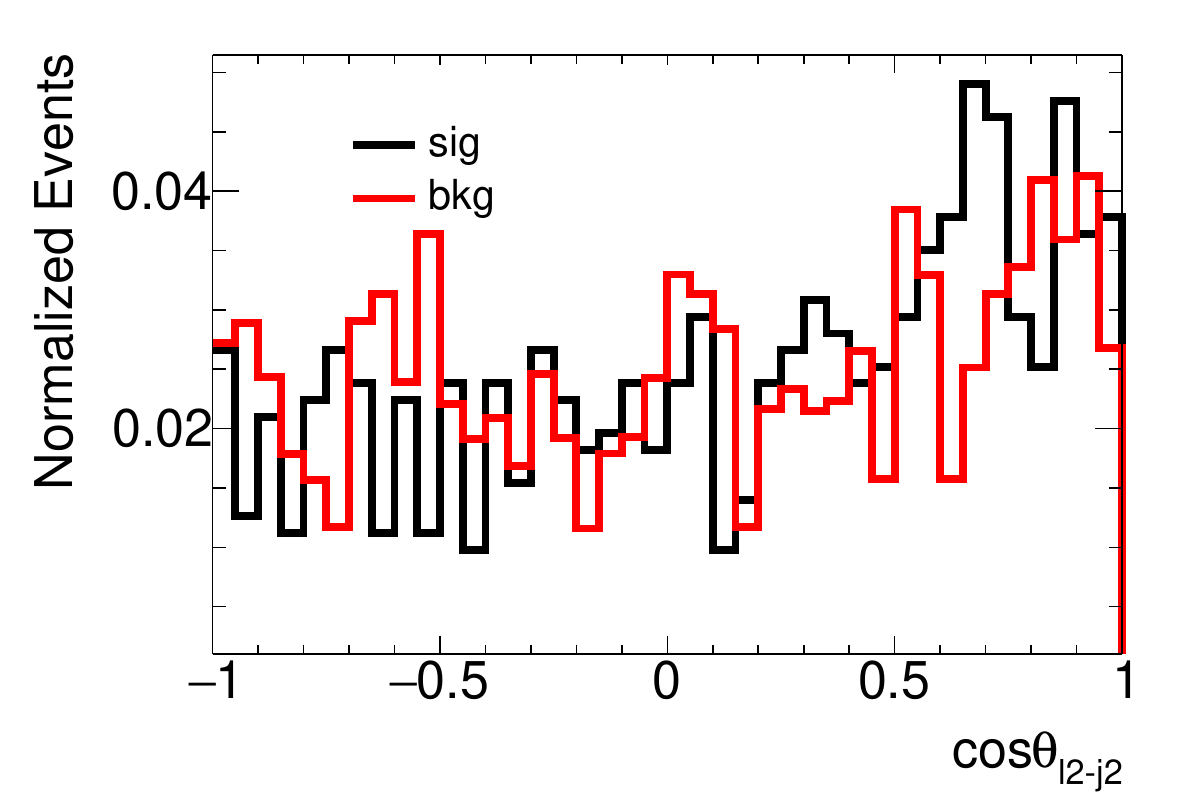}	
      \end{center}
\end{minipage}

\begin{minipage}{0.47\textwidth}
      \begin{center} 
                     (e)	\\
         \includegraphics[height=5.0cm]{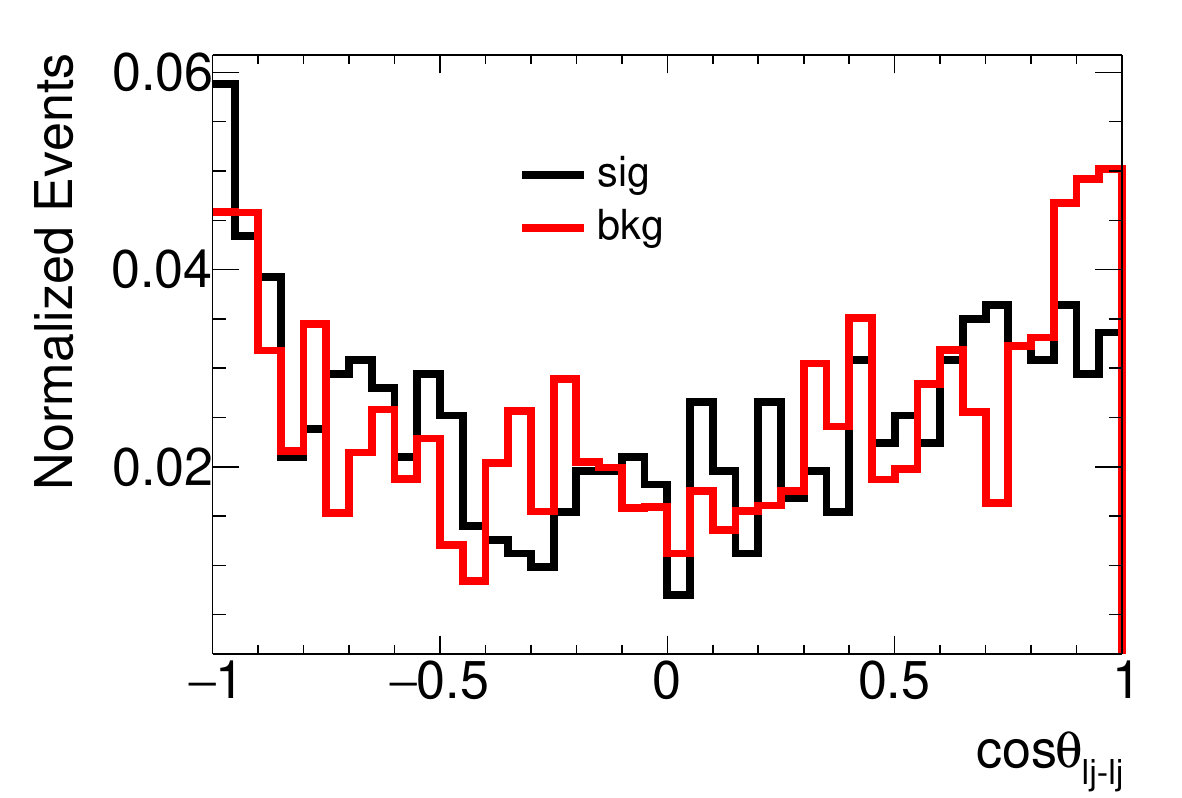}	
      \end{center}
\end{minipage}

 \caption{Normalised distributions in the (cosine of the) angle between  pairs of jets/leptons  (a)-(d), and two reconstructed bound state $(lj)^{1}$ and $(lj)^{2}$ (e) for BP4 and backgrounds at the LHC with $\cmsfourteen$  and $L~=~\threefbm$. 
 }\label{f:angle}
\end{figure}

A total of ten input variables are used in such a GBDT/TMVA analysis, as listed in Tab.~\ref{t:MVA_observables}. In addition to the invariant masses mentioned above, $\mttwo$, and $H_T$, we also compute five angular variables between pairs of final-state objects: $\cos(\theta_{l_1j_1})$, $\cos(\theta_{l_1j_2})$, $\cos(\theta_{l_2j_1})$, $\cos(\theta_{l_2j_2})$, and $\cos(\theta_{lj-lj})$. These angles are defined as the cosine of the angle between the two final states leptons $l$ and jets $j$, ordered in $E_T$, and $\cos(\theta_{lj-lj})$ is the angle between two reconstructed bound states $(lj)^{1}$ and $(lj)^{2}$. The distributions are shown in Fig.~\ref{f:angle}. As seen here, these variables are expected to be particularly useful since, in the signal process, they tend to be collinear due to the underlying boost structure, unlike in the background case.

\begin{table}[H]
 \begin{center}
 \begin{small}
\centerline{}
\begin{tabular}{|c|c|c|c|c|c|c|}
\hline
Energy variables& $M_{\text{lj}}^{1}$& $M_{\text{lj}}^{2}$ & $M_{\text{ll}}$ & $M_{\text{jj}}$ & $\HT$ & $\mttwo$\\
\hline \hline
Angular variables &$\cos(\theta_{lj-lj})$ & $\cos(\theta_{l_1j_1})$ & $\cos(\theta_{l_1j_2})$ &$\cos(\theta_{l_2j_1})$ & $\cos(\theta_{l_2j_2})$ &~ \\
\hline
\end{tabular}
 \end{small}
   \caption{The input observables used in the GBDT analysis.}\label{t:MVA_observables} 
 \end{center}
  \end{table} 

%

\begin{figure}[H]
	\centering
		\centering
		\includegraphics[width=8cm]{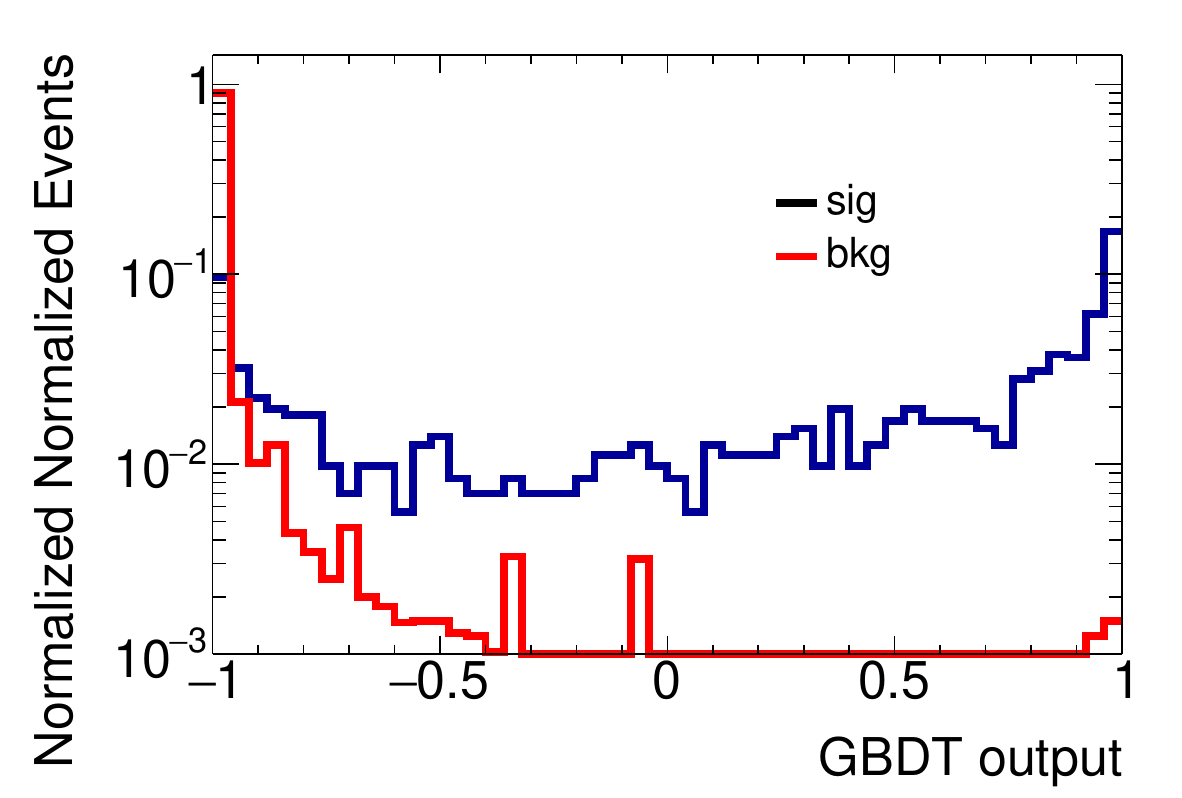}
\caption{GBDT output for BP4 and backgrounds. Label: signal in black colour, but the distribution is plotted in blue}\label{f:MVA}
\end{figure}

In the training stage, we take the signal and $ZZ$ background events as input. Because their physical process are very similar and difficult to distinguish apart in the selection with kinematic cuts.
The $\mttwo$ and the invariant mass $M_{\text{lj}}^{1,2}$ are the most important variables while training. But the angular variables  $\cos(\theta_{lj-lj})$ and  $\cos(\theta_{l_i j_k}),~i \neq k$ are also useful. Ultimately, for the GBDT output, a very good separation between the signal and background can be achieved, which is shown in Fig.~\ref{f:MVA}.

In the training stage, we take the signal and $ZZ$ background events as input, because their physical processes are very similar and difficult to distinguish in the selection with kinematic cuts.  
The $\mttwo$ and the invariant mass $M_{lj}^{1,2}$ are the most important variables during training, but the angular variables $\cos(\theta_{lj-lj})$ and $\cos(\theta_{l_i j_k}),~i \neq k$ are also useful. Ultimately, for the GBDT output, a very good separation between the signal and background can be achieved, which is shown in Fig.~\ref{f:MVA}.

\subsection{Significances}

The process of searching for signals at high-energy colliders using ML techniques is ultimately a statistical process. Therefore, to look for a signal, a large number of events need to be collected, and the kinematic variables of the final-state particles need to be analysed to find differences from the noise, but then a sizeable statistical sample needs to be collected, so as to have enough significance to claim sensitivity to the signal being searched for.  

The latter is defined here as $\Sigma=\frac{N_{\text{S}}}{\sqrt{(N_{\text{S}} + N_{\text{B}})}}$, where $N_{\text{S}}$ and $N_{\text{B}}$ represent the number of events for the signal ($\text{S}$) and (total) background ($\text{B}$), respectively, at a given luminosity $L$.

To distinguish the signal from the backgrounds and to compute the ensuing significance, we apply six kinematic cuts on reconstructed observables and one cut on the GBDT output, the details of which are listed in Tab.~\ref{t:BP4} (for BP4 in the signal case), where $L~=~\threefbm$ (with a centre-of-mass energy of 14 TeV, as usual). Following the acceptance cuts and upon applying the SS lepton criterion (the combination of which we refer to as preselection hereafter), we see a strong decrease in the initially huge SM backgrounds to the same order as the signal. For example, the troublesome $Zjj$ is almost eliminated after this cut (thus, it is no longer included in this table). As Figs.~\ref{f:Mh1Mh2}--\ref{f:MHHT} show, the kinematic selection is already quite efficient in separating the signal from the background, especially for the variable $\mttwo$.  
Finally, we also applied a GBDT discrimination before computing the final significances.

\begin{table}[H]
	\begin{center}
 \begin{tabular}{|c| c| c| c| c| c| c| c|}
 \hline
 $                  L~=~\threefbm  $ &  $S$ &  Top quark &  $W^\pm$  boson & $Z$ boson &     Total background &           $S/B$ &                          $\Sigma$\\ 
 
 \hline
Acceptance &                    938.2 &              $1.10\times 10^{6}$ &                  60544.8 &                   2584.9 &               $1.16\times 10^{6}$ &                      0.008 &                     0.87\\    
\hline
  SS Leptons &                    514.8 &                   2970.5 &                    301.4 &                    698.6 &                   3970.4 &                     0.130 &                                      7.69\\
\hline
 $                      \HT \in [ 60 ,400 ]~\gev$  &                    492.8 &                   2743.7 &                    260.6 &                    682.3 &                   3686.5 &                                        0.134 &                     7.62\\
\hline
$                  M_{\text{ll}} \in [ 10 ,300 ]~\gev $ &                    490.7 &                   2548.3 &                      235 &                    675.7 &                     3459 &                                        0.142 &                     7.81\\
\hline
$                M_{\text{lj}}^{1} \in [ 0 ,120 ]~\gev $ &                    434.8 &                   1656.5 &                    139.9 &                    486.5 &                   2282.9 &                                         0.190 &                     8.34\\
\hline
$                M_{\text{lj}}^{2}  \in [ 0 ,150 ]~\gev $ &                    415.6 &                   1490.1 &                    123.1 &                    457.6 &                   2070.8 &                                         0.20 &                     8.34\\
\hline
$                   M_{\text{vis}} \in [ 80 ,500 ]~\gev $ &                    364.7 &                    950.5 &                     88.7 &                    393.1 &                   1432.3 &                                         0.255 &                      8.6\\
\hline
  $                  \mttwo \in [ 10 ,110 ]~\gev $ &                    293.8 &                        0 &                        0 &                    149.5 &                    149.5 &                     1.965 &                         13.96\\ 
 \hline
  $                      {\rm GBDT} \in [ -0.35 ,1 ] $  &                    247.8 &                        0 &                        0 &                     56.8 &                     56.8 &                      4.363 &                          14.2\\
 \hline
 \end{tabular}
 \caption{Response to our selection cuts for the signal  (BP4) and background (separately and total) rates computed at the LHC with $\cmsfourteen$  and $L~=~\threefbm$.}\label{t:BP4} 
 \end{center}
 \end{table}

The results on $\Sigma$ for all 9 BPs are shown in Tab.~\ref{t:significance}. With only the acceptance and SS lepton criteria, the significances can reach up to 3. After applying the kinematic cuts, including exploiting the $\mttwo$ observable, the significances can be even larger than 10. Upon finally applying the GBDT selection, $\Sigma$ values can increase even further (albeit only slightly). In short, it is clear that a judicious combination of preselection and kinematic cuts, combined with a GBDT analysis, can result in large sensitivity to a significant expanse of the 2HDM Type-X parameter space through our chosen signal at the current LHC stage.

\begin{table}[H]
	\begin{center}
			\begin{tabular}{|c| c| c| c| c| c| c| c| c| c|}
				\hline
			     $\Sigma$        & BP1  & BP2   & BP3   & BP4  & BP5 & BP6 & BP7 & BP8 & BP9 \\
                    \hline
    After selecting SS leptons    & 6.22 &  5.11 & 7.71  & 7.69  & 5.94 & 6.42  & 5.88 & 8.08 & 5.4\\
                    \hline
    After kinematic cuts, w/o $\mttwo$ \& GBDT & 6.75 &  5.94 & 9.01  & 8.6 & 6.72 & 6.75 & 6.48  & 9.03 & 6.21\\
				\hline
    With $\mttwo$, w/o GBDT   & 11.76 &  10.81 & 14.38  & 13.96 & 11.54 & 11.33 & 11.37 & 14.65 & 11.24\\
				\hline 
    With $\mttwo$ \& GBDT     & 12.2 &  11.0 & 14.62  & 14.20 & 12.26 & 12.0 & 11.80 & 14.66 & 11.66\\
				\hline 
			\end{tabular}
			\caption{Significances following our different selections for all signals (BP1-9) at the LHC with $\cmsfourteen$ and $L~=~\threefbm$.
            }\label{t:significance}
	\end{center}
\end{table}


In order to have an overview on the scanned parameter space of the 2HDM, it is necessary to extend our analysis to the whole parameter space in our scan. To do such a work, it is quite demanding for computing resources which is beyond our reach. Instead, we introduce an approximate method to estimate the significance of each points in the parameter space. 

In our approximate method, we use the theoretical cross sections and the detector simulation efficiencies for signal events of each point to estimate the efficiencies. We use madgraph to compute the theoretical cross sections $\sigma_{theo}$. And we utilize the results of nine benchmark points presented above to fit the detector acceptance efficiencies. Acceptance efficiencies are defined as a ratio, labeled as $\epsilon_{acc}=\sigma_{\text{sim}}/\sigma_{\text{theo}}$, where $\sigma_{sim}$ is defined as the cross section after imposing all cuts. It is found that the acceptance efficiencies is proportional to $\mh + \mA$. Physically, it reflects that heavier Higgs bosons produce harder tau leptons, which are more likely to pass the lepton and jet pt cuts. In the parameter space of our study, it is found that the typical acceptance values range from 24\% to 33\%, so we take the averaged value i.e. $27.5\%$, in our method. Thus we obtain the estimated signal cross section which is defined as $\sigma_{sim} = \sigma_{theo} \times \epsilon_{acc}$. 

The detection efficiency describes the probability that produced particles can be  correctly identified and reconstructed. By fitting the benchmark point data, it is found that the jet detection efficiency is relatively stable, which can reach to 92.3\% or so.  However, the lepton detection efficiency shows a slight linear dependence on the Higgs boson masses, which simply reflect the fact that heavier particles produce more energetic tau leptons. A more energetic tau lepton can sequentially decay to an energetic lepton, which is easier for detectors to detect and reconstruct, thus a better lepton detection efficiency can yield. In the mass range of this study, it is found that the lepton detection efficiency can vary from 25\% to 29\%.

In order to improve signal selection and to suppress backgrounds, it is necessary to undergo the kinematic analysis and event pre-selection.  The efficiency of kinematic cuts is also extracted from the benchmark points presented above, and it is observed that the efficiency is negatively correlated with the total mass $\mh + \mA$ for the signal events. Using the data of all BPs, after all kinematic cuts, the dominant SM background events are from $ZZ$ process. Finally, we obtain all the fitted efficiencies, which are summarized as given below:

\begin{eqnarray}
    \epsilon_{acc} &=& 1.4764 \times 10^{-3} (\mh + \mA) - 0.0373, \nonumber \\
    \epsilon_{lep} &=& 8.66 \times 10^{-4} \times \mh + 0.1830, \nonumber \\
    \epsilon_{kin, sig} &=& -3.86 \times 10^{-4} \times (\mh + \mA) + 0.303, \nonumber \\
    \epsilon_{kin, ZZ} &=& 1.53 \times 10^{-3} \times (\mh + \mA) - 1.39 \times 10^{-5}  \mh  \mA - 0.148.    
\end{eqnarray}

\begin{figure}[H]
	\centering
		\centering
		\includegraphics[width=8cm]{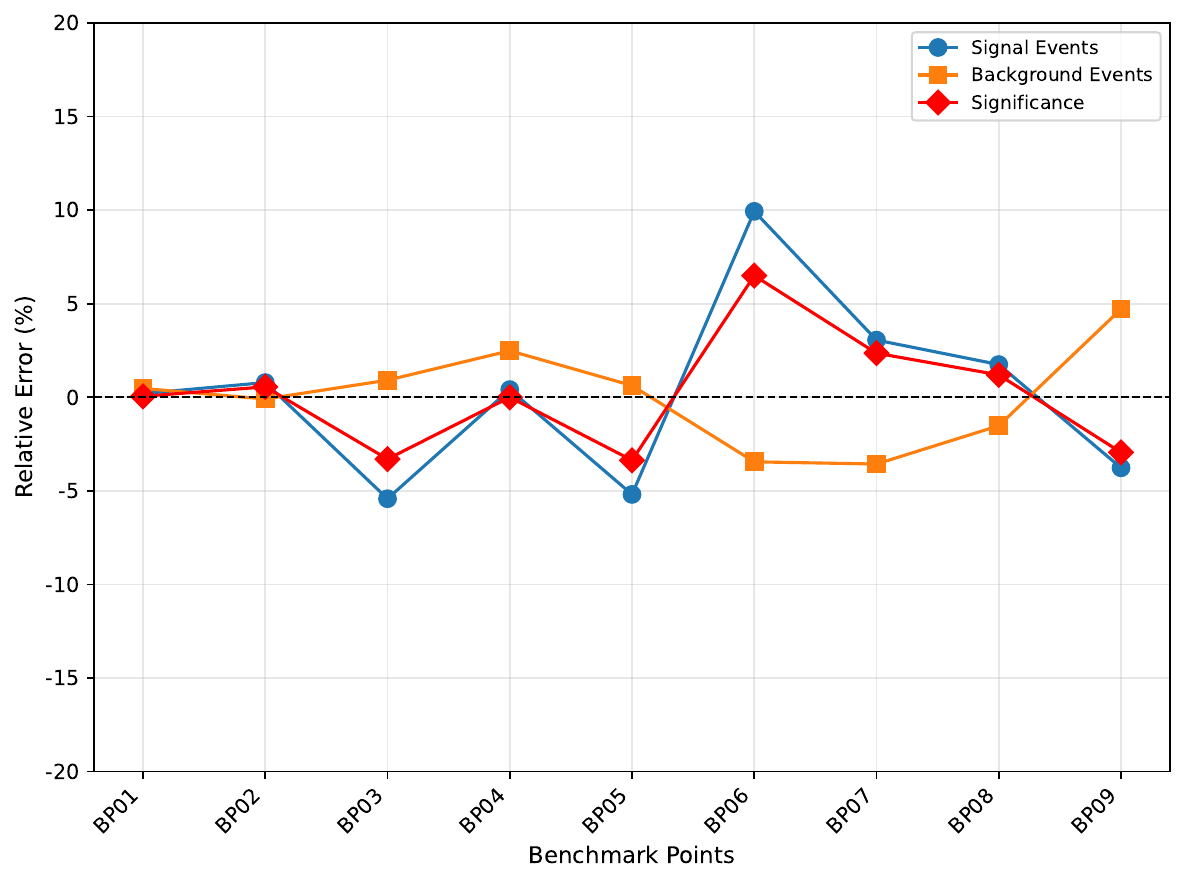}
\caption{The fluctuation of efficiencies are demonstrated for nine BPs.}\label{f:effi_estimation}
\end{figure}

Here we would like to comment on the errors in this fitting. The dominant error arises from fitting the kinematic cut region of the signal events, and the results are dependent upon the sample events of signal and background events. Even though, with our data samples, the current results yield at most a $6\%$ discrepancy in the obtained significance, as shown in Fig. \ref{f:effi_estimation}. In Fig. ~\ref{f:effi_estimation}, we show the fluctuation of significances in our approximate method when compared with the ones from the full simulation for the nine BPs. From the results, we can conclude that our approximate method works pretty well indeed. Thus we will apply our approximate method for a rapid estimates across a large set of parameter points. The final detectable cross section can be reliably expressed as $\sigma_{detectable} = \sigma_{sim} \times \epsilon_{jet}\times \epsilon_{lep}  \times \epsilon_{kin} $. 

\begin{figure}[!h]
	\centering
\begin{minipage}{0.47\textwidth}
      \begin{center} 
         \includegraphics[height=5.0cm]{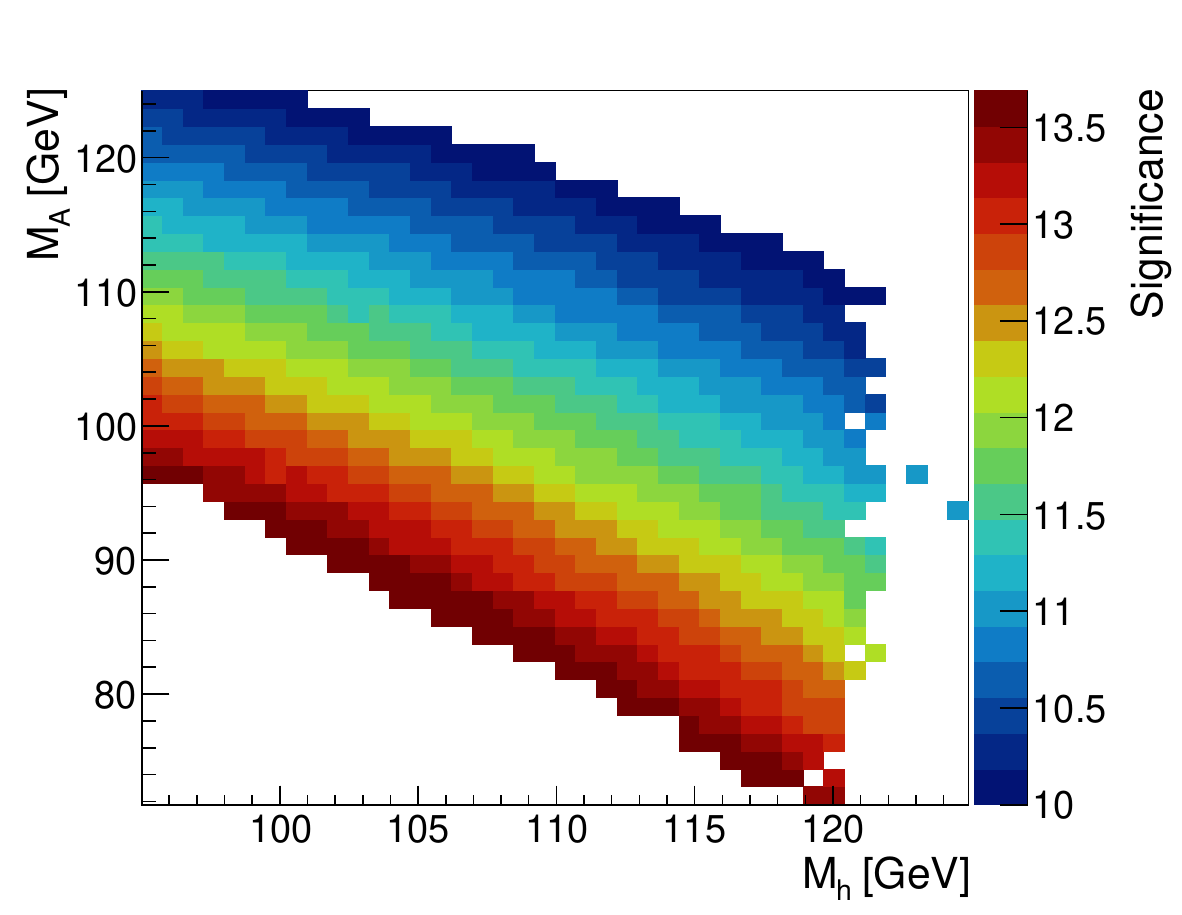}	
      \end{center}
\end{minipage}
\begin{minipage}{0.47\textwidth}
      \begin{center} 
         \includegraphics[height=5.0cm]{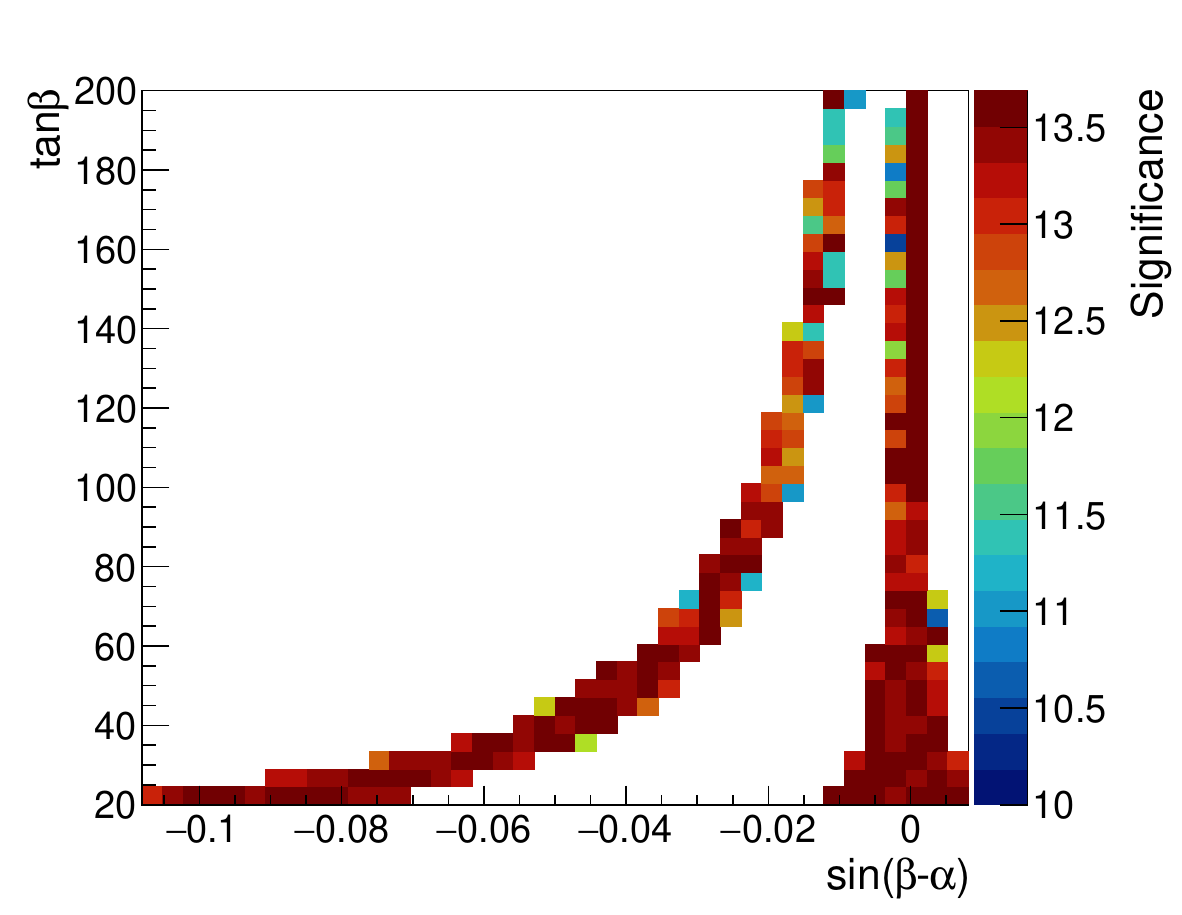}	
      \end{center}
\end{minipage}
\caption{Maximum significances for each grid in $\mh$-$\mA$ and $\sinbma$-$\tanb$ plane.}\label{f:heatmap_mh_mA}
\end{figure}

 The results for the points of parameter space are shown in Fig. ~\ref{f:heatmap_mh_mA}, where only maximum significances are shown in the each grid. It is observed that there exists a broad parameter space where the signal can be reachable at the LHC and the significance can reach up to 12 or higher. 

Furthermore, in order to display the distribution of significance in parameter space transparently, we demonstrate our results in the form of heatmaps. In the left and right plots, we divide the parameter space into grids of $\mh$-$\mA$ and $\text{sin}(\beta-\alpha)$-$\text{tan}\beta$, respectively. For each grid cell, we take the maximum significance among allowed points as the representative value, which is shown in Fig.~\ref{f:heatmap_mh_mA}. It is obvious that these points have a high potential to be detected at the LHC.

\section{Discussion and Conclusions}
In this paper, we have investigated the feasibility of hadro-producing the light Higgs scalar $h$ and pseudoscalar $A$ through $Z^{*}$ exchange (initiated by $q\bar q$ pairs) at the LHC in the 2HDM Type-X (or lepton-specific). We have focused on the $4\tau$ final state, where two $\tau$'s decay leptonically to form a pair of SS leptons (electrons and/or muons), and two $\tau$'s decay hadronically (thus producing two jets). In order to motivate future studies of this signature, we have also proposed 9 BPs and processed them through a realistic MC analysis for a centre-of-mass energy of $14$ TeV and (integrated) luminosity $L=\threefbm$, thus emulating actual LHC conditions. We have proven that this kind of signal could be found (or excluded) at the end of Run 3 of the LHC, following a dedicated selection based on both a kinematic and GBDT/TMVA analysis.

Although our study primarily focused on LHC sensitivity, the large $\tanb$ regions examined are also of interest for resolving the long-standing muon $g-2$ anomaly. BPs 5--9 with $\tanb$ larger than 100 could resolve this anomaly by enhancing one-loop ($\Hboson/\Aboson/\Hpm$) and two-loop (Barr-Zee) contributions. Future LHC data may narrow the viable explanations for the $g-2$ discrepancy. If the LHC excludes these high $\tanb$ regions, it would also constrain the 2HDM Type-X explanation for $g-2$.

In the sign/background analysis, we have deliberately avoid using the tau taggings to reduce the fluctuation of our analysis. Here let's address the effects of hadronic tau tagging briefly. Obviously, to introduce hadronic tau taggings would suppress both signal and background proportionally. For example, the sign mismeasured leptons from $Zjj$ cam be suppressed effectively, although such type of background can also be suppressed efficiently by the kinematic observables introduced in this work. Nevertheless, it is worthy of noting that after the pre-selection of same sign leptons, the dominant background is from $ZZ$ which can mimic our signal events mostly. Let's assume a work point with a hadronic tau tagging efficiency $0.6$. Then, to impose 1 haodronic tau tagging can reduce the significance by a factor $0.77$, while impose two hadronic tau taggings can reduce the significance by a factor of $0.6$. As shown in Tab. VII, even after taking two hadronic tau tagging, the discovery potential of BPs introduced in this work can be achievable at the LHC when the luminosity is sufficient.

\section*{Acknowledgments}
 SM is supported in part through the NExT Institute and the STFC Consolidated
Grant ST/X000583/1. SS is supported in full through the NExT Institute and
acknowledges the use of the IRIDIS High Performance Computing Facility, and associated
support services, at the University of Southampton, in the completion of this work. YW’s
work is supported by the Natural Science Foundation of China Grant No. 12275143, Central Guidance for
Local Science and Technology Development Fund Project No.2025ZY0020 and the Fundamental Research Funds
for the Inner Mongolia Normal University Grant No. 2022JBQN080. QSY is supported by the
Natural Science Foundation of China under the Grants No. 11875260 and No. 12275143.

\end{document}